\documentclass[12pt]{article}

\usepackage{graphicx,afterpage}  

\newcommand{\fig}[1]{Fig.~\ref{#1}}
\newcommand{\tabl}[1]{Table~\ref{#1}}
\newcommand{\eqr}[1]{Eq.~(\ref{#1})}

\begin{document}

\vspace{1cm}
\begin{center}

{\Large{\bf Topological structure and interaction}}\\

\smallskip

{\Large{\bf strengths in model food webs}}\\

\vspace{0.5cm}

{\it Christopher Quince$^{\rm a}$, Paul G. Higgs$^{\rm b}$ and Alan J.
McKane$^{\rm c}$}
\\
\bigskip

$^{\rm a}$Departments of Physics and Astronomy, Arizona State University, \\
P.O. Box 871504, Tempe, Arizona 85287-1504, USA \\

$^{\rm b}$Department of Physics and Astronomy, McMaster University, \\ 
Hamilton ON, Canada L8S 4M1

$^{\rm c}$Department of Theoretical Physics, University of Manchester, \\
Manchester M13 9PL, UK \\

\end{center}

\bigskip

\vspace{1cm}

\begin{abstract}
We report the results of carrying out a large number of simulations on a 
coevolutionary model of multispecies communities. A wide range of parameter 
values were investigated which allowed a rather complete picture of the 
change in behaviour of the model as these parameters were varied to be 
built up. Our main interest was in the nature of the community food webs
constructed via the simulations. We identify the range of parameter values 
which give rise to realistic food webs and give arguments which allow some
of the structure which is found to be understood in an intuitive way. Since
the webs are evolved according to the rules of the model, the strengths of 
the predator-prey links are not determined a priori, and emerge from the
process of constructing the web. We measure the distribution of these link
strengths, and find that there are a large number of weak links, in agreement
with recent suggestions. We also review some of the data on food webs 
available in the literature, and make some tentative comparisons with our 
results. The difficulties of making such comparisons and the possible 
future developments of the model are also briefly discussed. 
\end{abstract}

\vspace{0.5cm}


\newpage

\section{Introduction}

Food webs, which represent the links between predators and prey in an 
ecological community, are complex networks which present several problems 
to the modeller. Firstly, can the range and nature of webs be specified a 
priori, perhaps using some generic biological principles? Secondly, can a 
dynamics describing the change in population sizes of species in the web be 
defined on the network? Thirdly, given that the number of links in a food 
web of a reasonable size will be of the order of several hundred, is it 
possible in practice, or even desirable, to give values to all of these 
link strengths? 

The formidable difficulties associated with answering these questions led 
most early researchers to adopt approaches to the modelling of food webs 
which either bypassed some of these questions entirely, or implemented them 
in a very simple way. For example, May (1973) assumed the network to be a 
random graph, the interactions to be of randomly chosen strength and 
linearised the dynamics near a fixed point. Other food web modellers ignored 
the dynamics completely and simply gave rules to construct static 
graphs (Cohen et al., 1990; Williams and Martinez, 2000). Even 
in the recent flurry of interest concerning network structure and topology 
(Albert and Barab\'asi, 2002), most modellers have concentrated on 
specifying the nature of the network, rather than defining the dynamics on 
the network. The problem with this approach is that there is no reason 
why network structures which are appealing or which are found in social or 
other networks (small world, scale invariance) should apply to food webs. 
A recent study indeed shows that this is the case (Dunne et al., 2002a).

This suggests a more sophisticated approach to the modelling of food webs
should be adopted. A clue as to how we might move forward is that it is clear 
that the structure of the network depends on the dynamics of the network, and 
cannot be divorced from it. It is therefore not appropriate to specify 
the web and then define dynamics --- the two are interdependent. For
example, a predator-prey link between any two species will disappear if 
either of the species becomes extinct and this will depend on the nature of
the population dynamics that is chosen to govern their interaction. It is also
clear that the dynamics on the network will be strongly influenced by the 
nature of the network itself. This strongly suggests that we cannot separate 
the population dynamics on the network from the dynamics which changes the 
network structure, which will occur on much longer time scales.
 
These comments address the first two of the questions posed at the beginning 
of this section, but there still remains the difficulty of knowing how to 
generate the several hundred quantities which specify the parameters in the 
model dynamics at any given time. The solution to this problem which we 
favour is to assemble the food web from one (or very few) species, so that 
the parameters are determined by the dynamics of web assembly. We have 
already explained that this dynamics is inextricably linked to the population 
dynamics. Starting with only one species simply amounts to giving an initial
condition to the dynamics. In this way we determine those food web structures 
that can actually be reached rather than simply those that are possible. 

A model based on this philosophy was developed by some of us a few years ago 
and has been under study since then. An original version of the model 
(Caldarelli et al., 1998) was superseded by a later version with 
more realistic population dynamics (Drossel et al., 2001), and 
reviews which discuss the model specifically (Quince et al., 2002) 
and in a more general context (Drossel and McKane, 2003) are available. Our aim 
in this paper is to present a more extensive set of results from the model, 
emphasising aspects that were not stressed in previous investigations. A 
prime example is the distribution of link strengths, which is a topic which 
has been discussed extensively in the last year or so (Berlow et al., 
2004) and which is an emergent attribute in our model, and consequently a 
fundamental test of the whole approach.

We begin by outlining the model in Section 2 and then, in order to provide 
some intuition on how a particular web is built up, we describe the time 
evolution of a single web in Section 3. The structure of the model food 
webs which are dynamically constructed through simulation of the model 
are explored in Section 4 and compared with data in Section 5. The distribution
of link strengths in the model is explored in Section 6 and a discussion of 
our broad conclusions, as well as possible future avenues of investigation, 
is given in Section 7.   

\section{The model}

In this section we will give an overview of the model, presenting enough 
detail that subsequent sections of the paper may be understood. Readers 
should consult Drossel et al. (2001) for further details, especially 
regarding motivation for various model choices and the specifics of the
computer simulation.

The model is unusual in spanning a very large range of time scales, from 
changes in foraging strategies --- which might occur on a time scale of the 
order of days --- to evolutionary time scales. As we have indicated in the 
Introduction, we believe that phenomena on these different time scales cannot 
be separated, hence the need to consider them as a coherent whole.

On the shortest time scale the number of species and their populations are 
fixed, and only the foraging strategy --- the fraction of time that an 
individual of a particular species $i$ spends preying on individuals of 
another species $j$ --- changes. This fraction will be denoted by $f_{ij}$, 
and will be called the {\it effort} that species $i$ puts into preying on 
species $j$. Clearly $\sum_{j} f_{ij} = 1$ for all species $i$.

A reasonable foraging strategy would be one in which the amount of effort 
that a particular predator, $i$, put into preying on each of its prey, 
indexed by $j$, would be proportional to the gain in resources. Since the
rate at which an individual of species $i$ consumes individuals of species 
$j$ is just $g_{ij}$, the functional response, this amounts to assuming that
$f_{ij} \propto g_{ij}$, for a given species $i$ and all its prey 
species $j$. Using $\sum_{j} f_{ij} = 1$, gives
\begin{equation}
f_{ij} (t) = \frac{g_{ij} (t)}{\sum_{k} g_{ik} (t)}\,.
\label{ESS}
\end{equation}
The justification of the choice (\ref{ESS}) is discussed in greater depth by
Drossel et al. (2001) where it is shown that in the context of this 
model this is an evolutionarily stable strategy. It is also interesting that 
it fixes the efforts in terms of the functional response, which is a far 
more familiar quantity to ecological modellers, and which will be discussed 
later in this section. 

On a larger time scale, the number of species in the food web is still fixed, 
but the populations of these species changes, as well as their efforts. This 
is the regime of conventional population dynamics. In the model it is 
described by a balance equation for the rate of change of the number of 
individuals of species $i$ in the food web at time $t$, $N_{i} (t)$:
\begin{equation}
\frac{dN_{i}(t)}{dt} = \lambda\sum_{j}N_{i}(t)g_{ij}(t) - 
\sum_{j} N_{j}(t)g_{ji}(t) - d_{i}N_{i}(t)\,.  
\label{balance}
\end{equation}
Here $\lambda$ is the fraction of prey resource which is turned into 
predator births and $d_{i}$ is the constant rate of death of individuals 
of species $i$, in the absence of interactions with other species. 
These equations are the natural generalisations of balance equations found 
in the literature for systems with only a few species (Maynard Smith, 1974; 
Roughgarden, 1979), with the first term on the right-hand side representing
the growth in numbers of species $i$ due to predation on other species, the 
second term the decrease in numbers due to predation by other species, and 
the last term the death rate of individuals of species $i$. Where there is no 
predator-prey relationship between species $i$ and species $j$, $g_{ij}$ 
is zero. 

Finally, on still longer timescales, the number of species may change as 
well as their populations and efforts. Here we leave the realms of 
conventional population dynamics, and we need to give the species traits 
or {\it features} which define their behavioural and phenotypic 
characteristics. This will allow us to set up a scheme in which close 
variants of existing species are introduced into the community (speciation) 
and determine how good one species, $i$, is at preying on another, $j$ (the 
score $S_{ij}$). We do this by constructing a set of $K$ distinct features 
and an antisymmetric $K \times K$ matrix $m_{\alpha \beta}$ which gives 
the score of feature $\alpha$ against feature $\beta$. A new matrix is 
chosen at the start of every simulation run with entries which are Gaussian 
random variables with zero mean and unit variance. Species are then defined 
to be sets of $L$ distinct features. In the simulations we describe in 
this paper we took $L=10$ and $K=500$, but any two integers which allow for 
a very large number of distinct species to be created would be acceptable. 
The score of species $i$ against species $j$ is defined in terms of the 
scores of all the features of species $i$ against all the features of 
species $j$:
\begin{equation}
S_{ij} = {\rm max} \left\{ 0,\,\frac{1}{L}\sum_{\alpha \in i} 
\sum_{\beta \in j} m_{\alpha \beta} \right\}\,.
\label{scores}
\end{equation}
Note that $S_{ij} \geq 0$ and if $S_{ij} > 0$ then $i$ is adapted to prey on 
$j$. In addition, a species 0, representing the environment, is introduced 
at the start of a simulation, and is left unchanged throughout that 
particular run.

The dynamics on the largest, evolutionary, time scale can now be described. 
Once the population dynamics defined by (\ref{balance}) settles down to a 
new equilibrium value, a speciation is permitted to occur. This consists of 
choosing a species in the food web at random to be the parent species. One 
of its features is then randomly selected and changed randomly to another 
feature. This resulting child species is then introduced into the food web 
with a population of $N^{\rm child}$ and the parent population reduced 
by $N^{\rm child}$. In simulations reported in this paper we always take 
$N^{\rm child} = 1$. After the speciation has been carried out, 
Eq.~(\ref{balance}) is then integrated forward and the population dynamics 
are allowed to determine whether the population of the new species, the 
parent species, and indeed all other species in the web, grow or decline as 
a result of the change in composition of the web. We set a minimum population 
level, $N^{\rm min}=1$. If the population of a species falls below this level 
at any point in the simulation, this species is assumed to be extinct and 
is removed from the system.

This describes the essence of the model. It remains to choose the functional 
response, $g_{ij} (t)$. Considerable care was taken in making a biologically 
realistic choice; a detailed account of the logic behind the choice is given 
in Drossel et al. (2001). To motivate it, let us first describe the 
form we use, but for a single predator $i$ feeding on a single prey $j$:
\begin{equation}
g_{ij} (t) = \frac{S_{ij} N_{j} (t)}{b N_{j} (t) + S_{ij} N_{i} (t)}\,,
\label{1p_1p}
\end{equation}
This is known as a ratio-dependent function response (Arditi and Ginsburg, 
1996), since $g_{ij}$ is a only a function of the ratio $N_{i}/N_{j}$. The 
functional response which we actually use has to apply to a general web, 
where a given species may have an arbitrary number of predators and prey. 
We therefore modify (\ref{1p_1p}) by (i) introducing the efforts (\ref{ESS}), 
and (ii) replacing the term $S_{ij} N_{i} (t)$ by a sum of terms over all the 
predators of $j$ (denoted by $k$ and which includes $i$). This will involve 
a function $\alpha_{ki}$ which describes predator competition and which we 
take to have the form:
\begin{equation}
\alpha_{ki} = c + (1-c)q_{ki}\,.
\label{c}
\end{equation}
Here $c$ is a constant such that $0 \leq c \leq 1$ and $q_{ki}$ is the 
fraction of features of species $k$ that are also possessed by species $i$.
The result is a generalised ratio-dependent functional response:
\begin{equation}
g_{ij} (t) = \frac{S_{ij} f_{ij} (t) N_{j} (t)}{b N_{j} (t) + 
\sum_{k} \alpha_{ki} S_{kj} f_{kj} (t) N_{k} (t)}\,.
\label{g_ij}
\end{equation}
The choice for the competition function $\alpha_{ki}$ (\ref{c}) is motivated
by the expectation that species which are different from each other (small 
$q_{ki}$) should be less in competition for resources (i.e. individuals of 
species $j$) than those which are similar.

A flow diagram showing the sequence of steps in a single simulation is shown 
in Fig.~1. The simulation starts with the choice of species 0 (the environment)
and one other species which will begin by feeding off it (an autotroph). From
a practical point of view, the differential equations (\ref{balance}), are
solved numerically by discretising time into segments of duration $\Delta T$. 
At the beginning of one of these periods the population numbers 
$\left\{ N_{i} (t) \right\}$ have just been updated, and therefore new 
functional responses may be determined from (\ref{g_ij}). The process of 
iterating (\ref{ESS}) and (\ref{g_ij}) to produce new efforts corresponding 
to these new populations take place on the shortest time scale of the model. 
The iteration of the population dynamics (\ref{balance}) takes place on 
the intermediate time scale, and the speciation process on the longest 
time scale. 


\begin{figure}[t]
\begin{center}
\rotatebox{0}{\scalebox{2.5}{\includegraphics[width=.3\textwidth]
{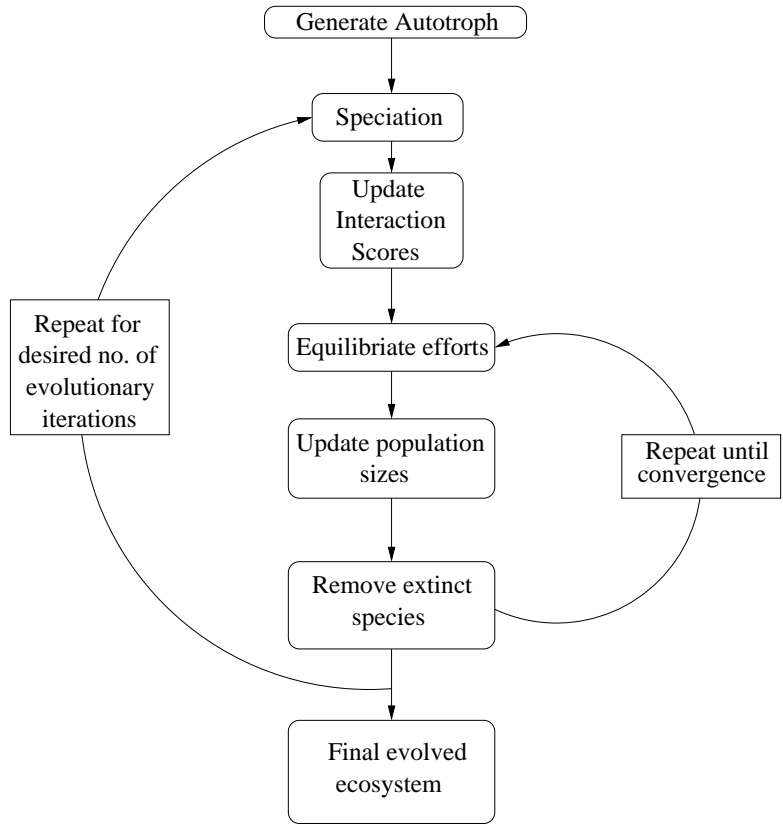}}}
\end{center}
\caption[A diagram summarising the model.]{Diagram illustrating the simulation process.} 
\label{summary}
\end{figure}


Finally, we have mentioned several parameters during the description of
the model, such as the parameter $b$ in the definition of the functional
response (\ref{g_ij}) and the competition parameter $c$ in (\ref{c}). The
other main parameter of the model is denoted by $R$, and is the rate of 
input of resources from the environment into the food web. These parameters,
as well as four others which are kept fixed throughout our investigation, 
are given in Table 1.


\begin{table}[t]
\begin{center}
\renewcommand{\arraystretch}{1.4}
\setlength\tabcolsep{5pt}
\begin{tabular}{lll}
\hline\noalign{\smallskip} Symbol & Name & Description\\
\noalign{\smallskip} \hline \noalign{\smallskip}
$R$ & Resources & Rate of input of external resources into food web\\
$c$ & Competition constant & Determines the degree of inter-specific 
competition\\
$b$ & Saturation constant & Controls the effectiveness of predation\\
\noalign{\smallskip} \hline\noalign{\smallskip}
$\lambda$ & Conversion efficiency & Ratio of numerical to functional 
responses\\
$d$ & Death rate & Per capita death rate in the absence of interactions\\
$N^{\rm min}$ & Minimum population & Population below which species are 
assumed extinct\\
$N^{\rm child}$ & Child population & Population at which new species are 
added\\
\hline
\end{tabular}
\end{center}
\caption[The parameters of the model.]{The parameters of the 
model. Only the first three parameters are varied in this paper.}
\label{parameters}
\end{table}


\section{Time evolution of an individual simulation}


\begin{figure}[t]
\begin{center}
\rotatebox{0}{\scalebox{2.5}{\includegraphics[width=.3\textwidth]{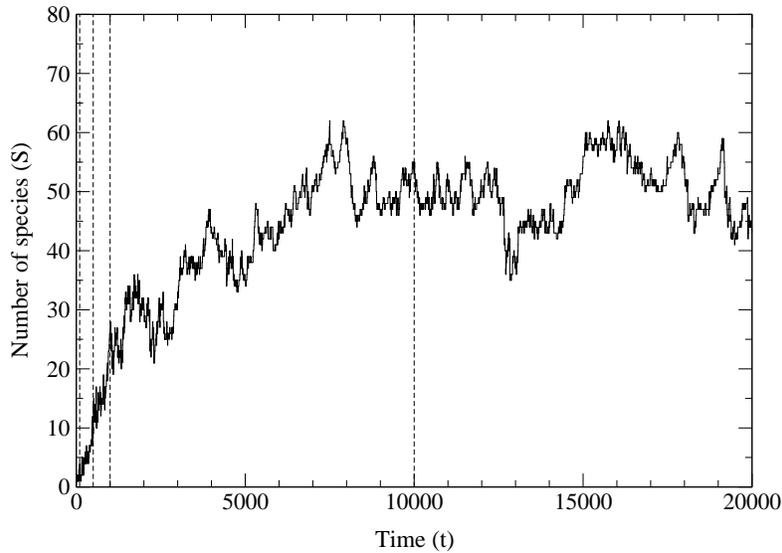}}}
\end{center}
\caption[]{The number of species, $S$, as a function of time, measured in 
number of speciation events for a single simulation of the model. 
This simulation had parameters $R = 1\times 10^5$, $b = 0.005$, $c = 0.5$. 
The dashed lines show the times 100, 500, 1000 and 10000. The food webs at 
these times are displayed in \fig{tfoodwebs}.} 
\label{series}
\end{figure}


The mechanism summarised in \fig{summary} is capable of generating large 
complex food webs. This is true for a wide range of parameter space --- the 
boundaries of which we explore in the next section. In \fig{series} the time 
evolution, measured in number of attempted speciation events, of the species 
number is shown for one of these sets of parameter values. For these values
of the parameters, the number of species initially increases quite rapidly 
but with sizable fluctuations. Then after about 10000 iterations it appears 
that the long time average number of species is approaching a constant value. 
There is considerable variation about this value representing a continuous 
overturn of species due to speciation and extinction.

Since speciation is represented as a random event in the model 
then the sequence of food webs generated during an individual model 
simulation can be viewed as a realisation of a stochastic process. For 
each particular choice of initial conditions corresponding to a set of 
parameter values, the randomly assigned feature matrix $m_{\alpha \beta}$ 
and random environment features, there will be a different time dependent 
probability distribution of food web configurations. The corresponding 
distribution for any food web statistic, such as the number of species 
shown in \fig{series}, may then be obtained from these configurations. That
the long time average number of species appears to become constant in 
\fig{series} suggests that it may be evolving towards a stationary 
distribution. This has been investigated further by examining extremely 
long simulations, of the order of $10^6$ speciation events. For these 
simulations no long term trends in any statistics were found after the 
initial growth phase. From this we conjecture that the probability 
distribution of food web structures also evolves towards a time 
independent stationary distribution.


\begin{figure}
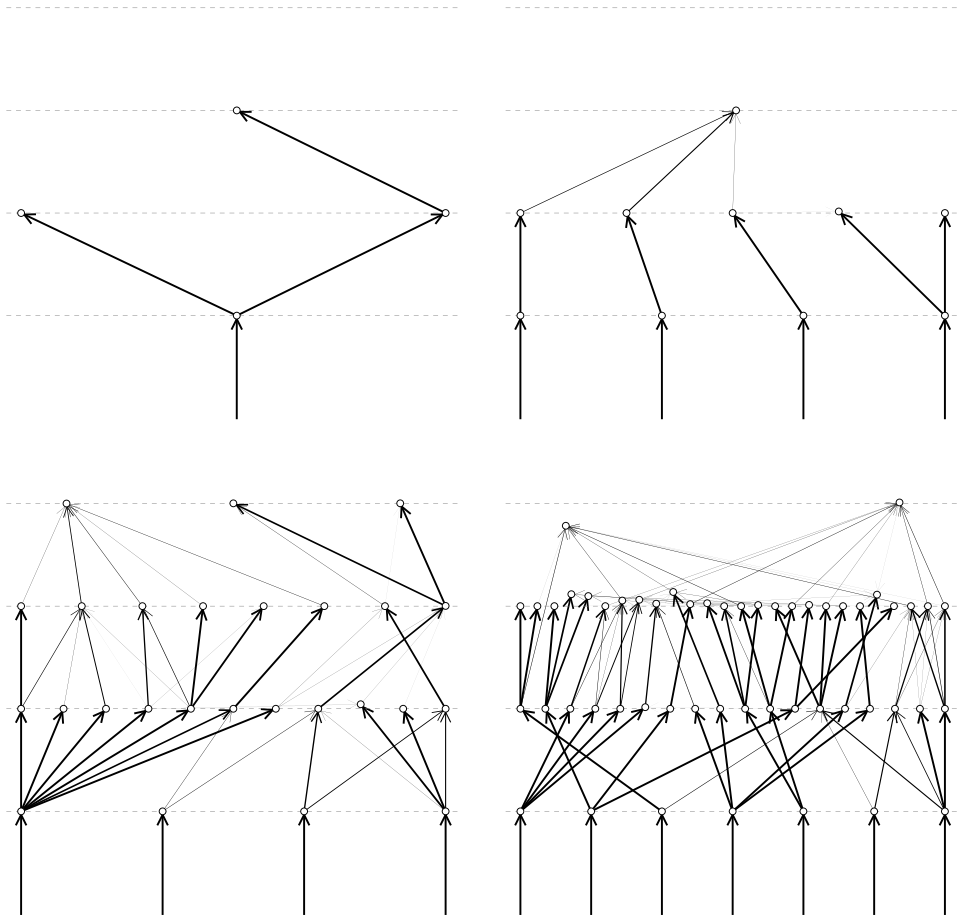

\begin{center}
\includegraphics*[width=6.0cm]{0results100.epsi} \hskip 5mm 
\includegraphics*[width=6.0cm]{0results500.epsi}
\vskip 10mm
\includegraphics*[width=6.0cm]{0results1000.epsi} \hskip 5mm 
\includegraphics*[width=6.0cm]{0results10000.epsi}
\end{center}
\caption[]{Four food webs sampled at times 100 (top left), 500 (top right), 
1000 (bottom left) and 10000 (bottom right) from the 
simulation described in the caption of \fig{series}.}
\label{tfoodwebs}
\end{figure}


The simulation shown in \fig{series} was sampled after 100, 200, 1000 and 
10000 speciation events and the food web structure at these times is shown 
in \fig{tfoodwebs}. The circles in these diagrams correspond to species and the
arrows represent predator-prey interactions between species. The arrows 
point from prey to predator and have thickness that is proportional to the 
fraction of the predator's diet constituted by the prey. The vertical arrows 
originating from the base of the diagrams indicate feeding off the 
environment. Only those links that constitute greater than 1\% of a 
predator's diet are shown in the diagram.

The horizontal position of the species in \fig{tfoodwebs} has no significance 
except to generate comprehensible diagrams. Species vertical position on the 
other hand does have meaning, it is proportional to what we will refer to as 
the \emph{trophic height} of the species. This is calculated as the 
weighted average of the lengths of the paths from the species to the 
environment, with the paths weighted multiplicatively by the predator 
diet fractions or efforts $f_{ij}$:
\begin{eqnarray}
h_{i} & = & 1 + \sum_{j = 1}^{S} f_{ij}h_{j}, \\
\Rightarrow \ \ h_{i} & = & \sum_{j = 1}^{S} (\delta_{ij} - f_{ij})^{-1}, 
\nonumber
\end{eqnarray}   
where $h_{i}$ is the trophic height of the $i$th species. The horizontal 
dashed lines in the food web diagrams have vertical position equal to 
integer trophic heights. We will also use the term \emph{trophic level} to 
indicate the minimum path length from a species to the environment and 
denote it by $l_{i}$. A majority of species in the model food webs have a 
trophic height close to their trophic level. Thus we will use the latter in 
this paper to investigate food web structures, since it has the advantage of 
being discrete and being defined for the typically binary empirical food webs. 
A recent review of the trophic level concept is given by Williams and 
Martinez (2004). Our definitions of trophic level and trophic height 
correspond to their shortest and flow based definitions of trophic level 
respectively. 

The diagrams in \fig{tfoodwebs} give a clear sense of the increase in 
average community complexity, in terms of species number and the number 
of predator-prey interactions per species, that occurs during the growth 
phase of the simulations. It is worth emphasising again that there are 
considerable fluctuations and that through extinction events the short term
trend can be a decrease in complexity. The quantitative statistics of the 
assembly process are investigated in Quince et al. (2002). The mature 
food web at time 10000 in \fig{tfoodwebs} shows some ecologically interesting 
features. A trophic structure has developed populated by a mixture of 
predators with many prey (generalists) and those with few prey (specialists). 
Most species exploit prey on the level below them but there are some omnivores.
There are also `functional groups' of species, a functional group being 
defined as a subset of species which share the same predators and prey 
(Walker, 1992). In the next section we will present statistics from these
long time structures.

\section{The structure of the model food webs}
\label{structure}

In the previous section it was argued that after a large enough number of 
speciation events the model generated food web structures from an individual 
simulation will be drawn from an approximately stationary distribution. It 
is this stationary distribution of structures that we are interested in here. 
Our aim is to provide a qualitative understanding of the structures
and the processes that generate them. This will be done by presenting, for 
a range of parameter values, both individual instances of food webs and 
descriptive statistics from ensembles. A similar analysis has been 
performed previously (Drossel et al., 2001), but the results here 
include a wider range of parameter values and have a different emphasis.

The stationary distribution of structures observed in any given simulation 
will depend not only on the model parameters themselves, but also on the 
particular realisations of the random feature matrix and environment 
species used. This was addressed by performing a dual averaging procedure 
to calculate the statistics in this section. They were first time averaged 
over the final part of each simulation to obtain approximations to the 
means of the stationary distributions and then ensemble averaged over multiple
runs, each with different realisations for the feature matrices and 
environment species. 

To perform a full investigation of the effect of the model parameters 
would require simultaneously altering all seven of them 
(see \tabl{parameters}). The computational effort involved in each individual 
simulation coupled with the need for multiple runs at each set of parameter 
values means that this is beyond the scope of this study. Instead we will 
restrict our attention to the main three parameters ($R$, $c$ and $b$) and 
ask how the food webs change as we alter each one of these parameters 
independently of the other two. The other four parameters will be kept 
constant in this section, and in fact throughout this paper,
with the values $\lambda = 0.1$, $N^{\rm min} = 1.0$, $N^{\rm child} = 1.0$ 
and $d = 1.0$. 


\begin{figure}[t]
\begin{center}
\rotatebox{0}{\scalebox{2.5}{\includegraphics[width=.3\textwidth]{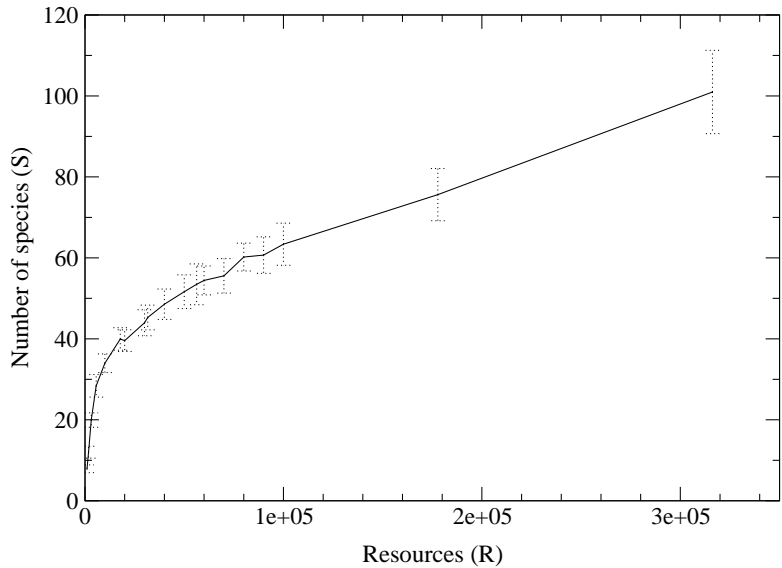}}}
\end{center}
\caption[]{The effect of the rate of input of resources on the number of 
species in the communities. The data points show ensemble averages over 
multiple simulations of the time averaged number of species. The simulation 
details are contained in the text. The error bars show standard deviations.} 
\label{rsfig}
\end{figure}


\subsection{Change in food-web structure with $R$}

We begin by considering the effect of altering the parameter $R$, the rate 
of resources input into the food webs. The data set for this study 
consisted of 20 independent simulations at each of 19 different values of 
$R$. The actual values used were $R = 10^{3+x}$, with $x=0.25n$ and 
$n=0,\ldots,10$, and $R=m \times 10^{4}$, with $m=2,\ldots,9$. An 
additional 60 runs with $R=100000$ were generated, making a total of 80 
for this particular value. The other two major parameters were kept constant 
with $b = 0.005$ and $c = 0.5$. The runs were independent in the sense that 
different sets of pseudo-random numbers were used in their generation. They 
therefore had different feature matrices, $m_{\alpha \beta}$, different 
environment species, and different speciation events throughout. All runs 
lasted for 120000 speciation events, except for $R=316200$, where the model 
was run for 220000 speciation events until the web ceased to grow on average. 
The statistics presented here were calculated by time averaging over the 
last 20000 iterations of each run and then ensemble averaging over 
all runs at a particular set of parameter values. 

In \fig{rsfig} we show the mean number of species as a function of $R$ 
for these simulations, from which it can be seen that as $R$ is increased 
the number of species in the ecological communities also increases. This 
is due to $R$ scaling the population sizes and hence allowing more species 
to exist with populations above the critical value ($N^{\rm min}$). In fact, 
the characteristic ratio $R/\lambda N^{\rm min}$ should control this effect.


\begin{figure}
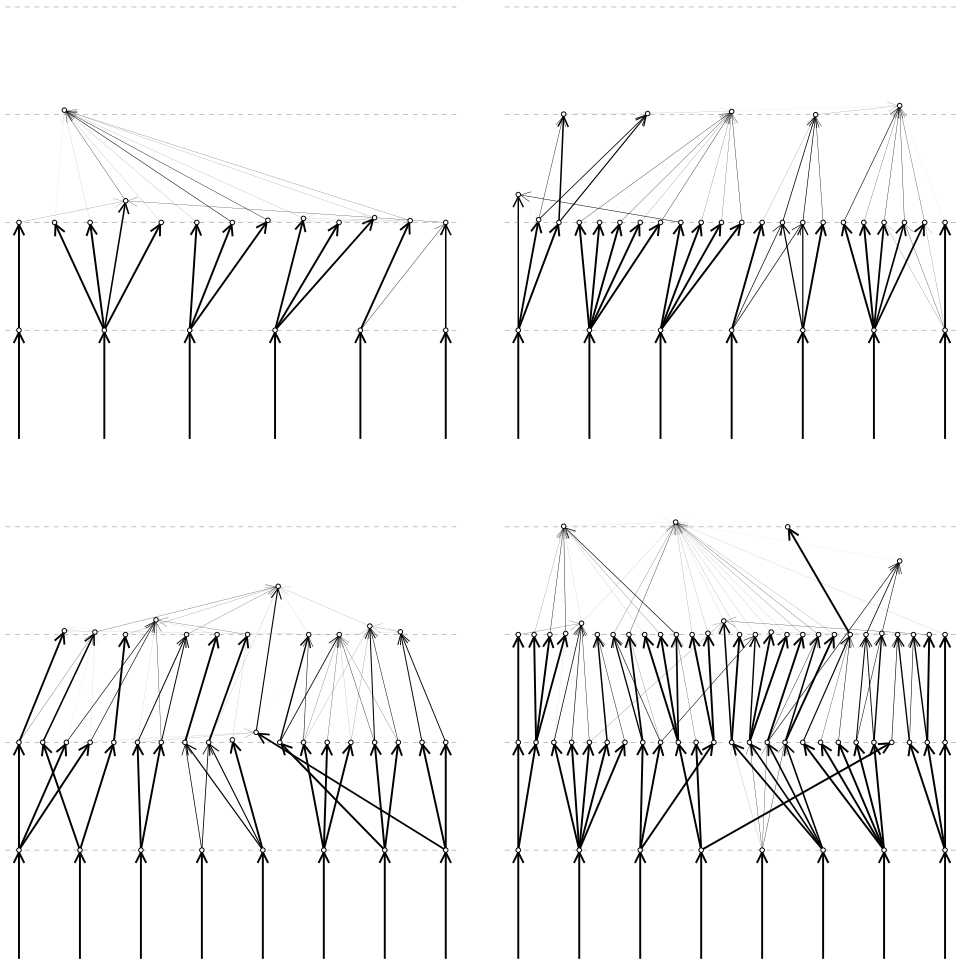

\begin{center}
\includegraphics*[width=6.0cm]{1e35.epsi} \hskip 5mm 
\includegraphics*[width=6.0cm]{1e4.epsi}
\vskip 10mm
\includegraphics*[width=6.0cm]{1e45.epsi} \hskip 5mm 
\includegraphics*[width=6.0cm]{1e5.epsi}
\end{center}
\caption[]{Four example food webs generated by the model. These 
food webs are the final states, after 120000 speciation events, from four 
of the simulations used to investigate the effect of altering $R$. The 
simulations had $R$ values of $10^{3.5}$ (top left), $10^{4}$ (top right), 
$10^{4.5}$ (bottom left) and $10^{5}$ (bottom right). All other parameters 
were kept constant as described in the text.}
\label{rfoodwebs}
\end{figure}


The parameter $R$ does not only impact on the species diversity of the food 
webs, it has considerable effect on the food web structure. This is 
illustrated graphically in \fig{rfoodwebs}, using the same conventions 
as \fig{tfoodwebs}, where four food webs with four different values of $R$, 
such that $R$ increases from left to right and from top to bottom in the 
figure, are displayed. 


\begin{figure}[t]
\begin{center}
\rotatebox{0}{\scalebox{2.5}{\includegraphics[width=.3\textwidth]
{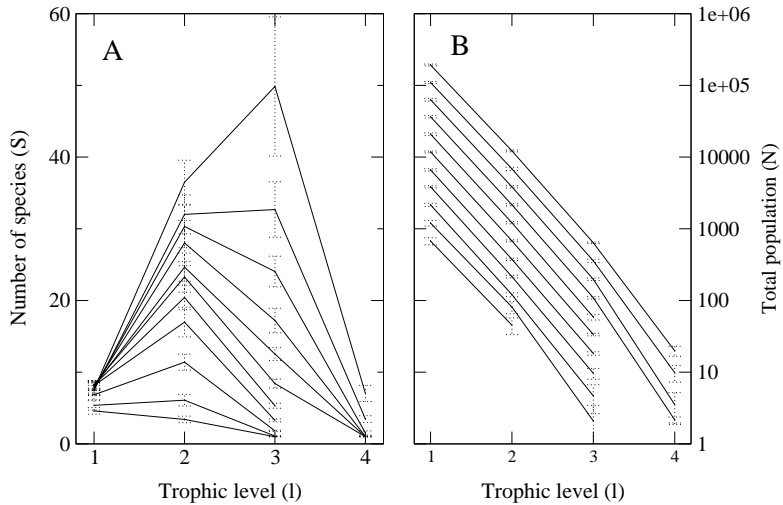}}}
\end{center}
\caption[]{The mean number of species (A) and total population (B) on each 
trophic level for simulations spanning eleven logarithmically scaled 
values of $R$. The actual simulations used were the subset of the 
simulations used to investigate the effect of altering $R$ described in 
the text with $R = 10^{3+x}$, $x=0.25n$ and $n=0, 1,\ldots,10$. The $R$ 
values of the lines can be determined by noting that species number on 
each level and total population always increases with $R$. The error 
bars in both graphs show standard deviations.} 
\label{rsplevelfig}
\end{figure}


In \fig{rsplevelfig}A the number of species occupying each trophic level 
averaged over multiple simulations at different $R$ values is shown. 
At a given value of $R$ the distributions are hump shaped indicating 
that the largest occupation numbers occur for intermediate trophic levels. 
As $R$ increases the number of species occupying all levels increases, 
but at varying rates, so that the peak of the distribution shifts to 
higher levels. The same hump shaped distribution of species between 
trophic levels and its dependence on $R$, or equivalent parameter, has 
been found both in the earlier version of the model (Caldarelli 
et al., 1998) and in a mean-field approximation to a Lotka-Volterra 
evolving food web model (L\"assig et al., 2001).

To the right of this figure the total population of each level ($N$) 
averaged over the same simulations is shown (\fig{rsplevelfig}B). The 
total population does not show a peak as for the occupation numbers. 
Instead it declines approximately geometrically with level at a particular 
$R$ value and on each level the total population is roughly proportional to 
$R$. The reason for this decline is that setting $\lambda$ to be smaller 
than one ensures that predator populations will in general be smaller 
than their prey populations.


\begin{figure}[t]
\begin{center}
\rotatebox{0}{\scalebox{2.5}{\includegraphics[width=.3\textwidth]
{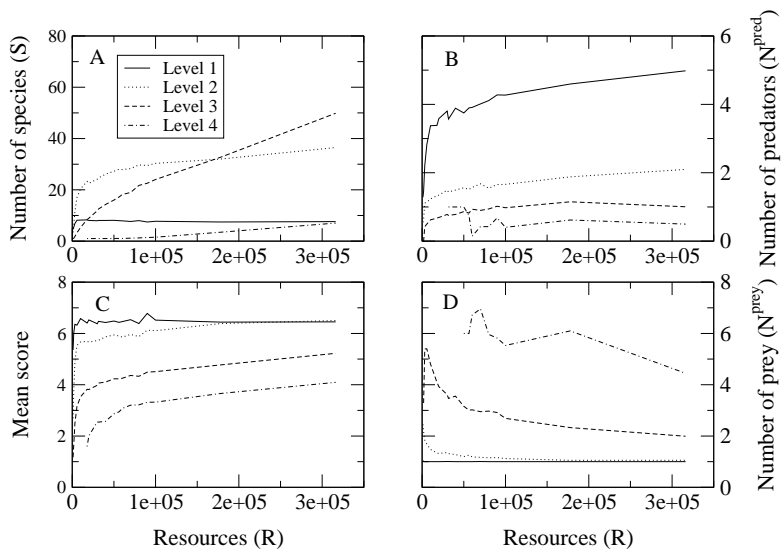}}}
\end{center}
\caption[]{The effect of $R$ on four quantities as a function of trophic 
level. A: the number of species on each trophic level. B: the number of 
predators averaged over all species on the same trophic level. C: the mean 
score averaged similarly. D: the number of prey again averaged over 
trophic level. The details of the simulations are contained in the text.} 
\label{r4levelfig}
\end{figure}


An intriguing feature of the four food webs shown in \fig{rfoodwebs} is 
that the species on higher trophic levels appear to be less specialised and 
exploit more prey, than those on the lower levels. In addition, the species 
on lower trophic levels have more predators on average than those on the 
higher levels. These patterns are also affected by $R$, so that as $R$ 
increases the species on a trophic level become more specialised and 
have more predators on average. That these patterns are general is confirmed 
by the statistics presented in \fig{r4levelfig}, where four different 
quantities have been averaged over all the species on a given trophic level 
and these quantities were, as usual, then both time averaged and ensemble 
averaged. The upper right hand graph (\fig{r4levelfig}B) and the lower 
right hand graph (\fig{r4levelfig}D) give the average number of predators 
($N^{\rm pred}$) and the average number of prey ($N^{\rm prey}$) per 
species for each trophic level as a function of $R$ respectively. In 
calculating the number of predators and prey all trophic links constituting 
greater than 1\% of a predators diet were counted and for the purposes 
of calculating the prey number the environment was treated as just 
another species. Thus it appears that predator number does indeed decrease 
with trophic level, and increase with $R$, whilst conversely prey number 
increases with trophic level and decreases with $R$.


\begin{figure}[t]
\begin{center}
\rotatebox{0}{\scalebox{1.5}{\includegraphics[width=.3\textwidth]{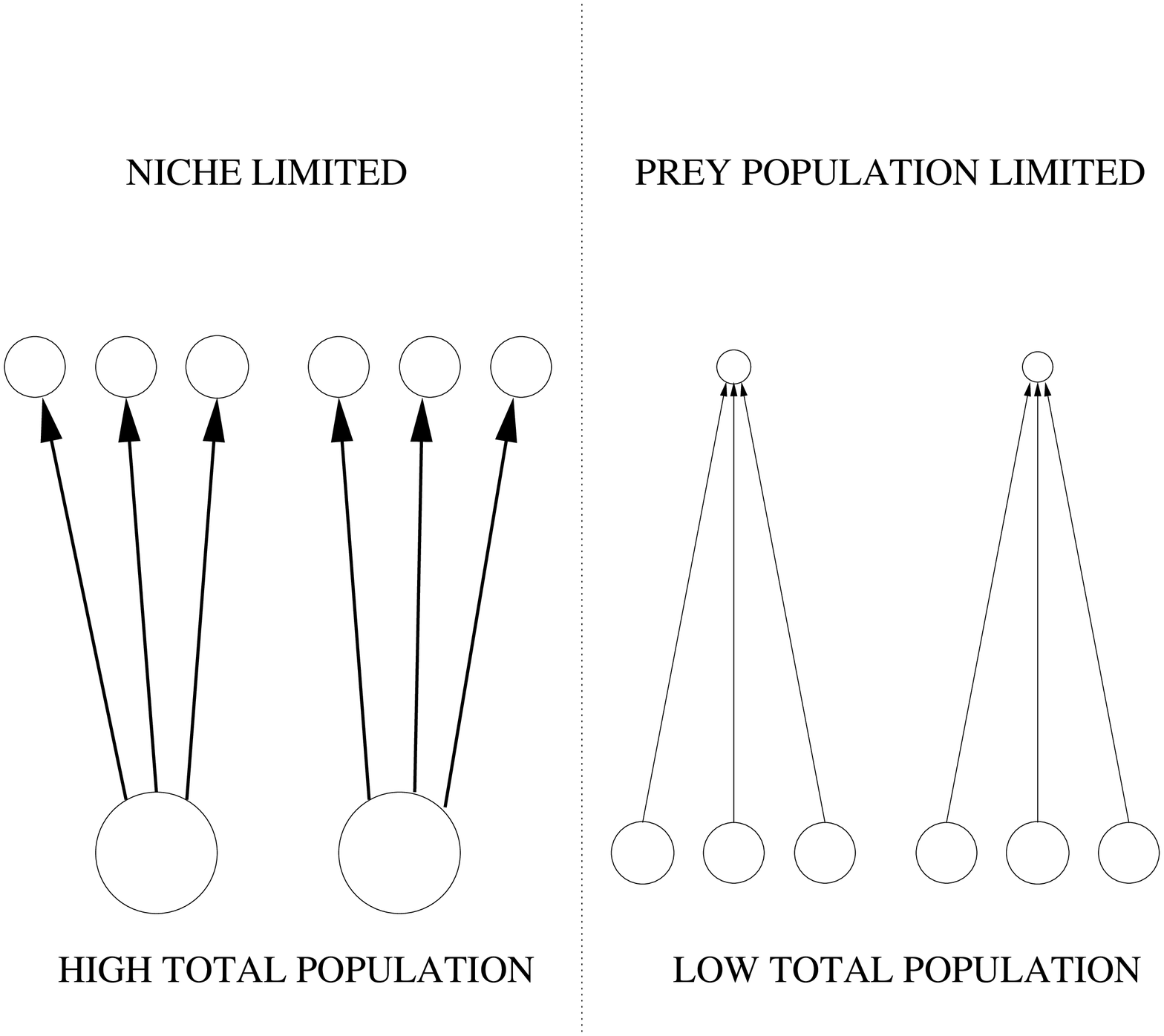}}}
\end{center}
\caption[]{A diagram illustrating the two hypothesised regimes of niche 
limitation and prey population limitation that control the species 
diversity of a trophic level.}
\label{Limit}
\end{figure}


These observations can be used to understand the observed patterns in 
trophic level occupation by hypothesising that each trophic level exists 
on a continuum lying between two regimes, its position on the continuum 
being determined by the total population of the level below. We will refer 
to the regime at the bottom end of this continuum as `prey population 
limited'. This regime is characterised by generalist species with very 
little prey overlap between competing species. The species diversity of 
the trophic level for this regime is limited by the total population of 
the level below. The other end of the continuum, when the prey population
on the level below is very high, we will designate the `niche limited' 
regime. Here species are specialised to feed off only one prey and many 
species exploit the same prey. In this regime the species diversity of 
the trophic level is limited by the total number of prey in the level 
below and the number of species that can exploit the same prey. A diagram 
illustrating the two regimes is shown in \fig{Limit}. 

The total population of a level decreases with trophic level (see 
\fig{rsplevelfig}B). Thus as we go up the levels we pass from the niche 
limited to the prey population limited regimes. For trophic levels near 
the niche limited regime, the number of available niches, and hence 
species diversity, increases with trophic level. This is because the 
number of species in the level below increases from just one, the 
environment, as trophic level increases. For trophic levels in the prey 
population limited regime the species diversity will decrease for higher 
trophic levels, since the population of the level below decreases. This 
then explains the humped shaped distributions of \fig{rsplevelfig}A. 
This argument owes a great deal to the ideas presented in L\"assig 
et al. (2001) and Bastolla et al. (2002).

The hypothesis also explains the changes in trophic level occupation 
with $R$ which are more easily understood from \fig{r4levelfig}A, where $R$ 
is plotted on the $x-$axis, than \fig{rsplevelfig}A. Initially at low $R$ 
values all trophic levels are prey population limited and their species 
diversity increases with $R$. The species diversity of the first trophic 
level then becomes niche limited, this occurs at very small $R$ values. 
Increasing $R$ further does not result in more level 1 species, since the 
maximum number of species that can exploit a single resource has been 
reached. The species diversity of the other three levels continues to
increase with $R$, until at very high $R$ values, level 2 becomes niche 
limited and its occupation saturates to a constant value. This value is 
much higher than that for level 1, since there are more prey available to 
level 2.

The above ideas are consistent with the observed changes in prey number 
with trophic level and $R$ (\fig{r4levelfig}A), but they do not explain 
why species specialisation should increase with the population of their 
prey. It seems likely that this occurs within the model because a predator 
lineage that predates only a few prey can evolve to be a better adapted 
predator than one that exploits many. This will be true because the 
$m_{\alpha\beta}$ determining the effectiveness of one feature with 
respect to another are independently distributed. Thus it is much less 
likely that changing one of a predator's features for another, as occurs 
in a speciation event, will simultaneously improve all the scores of a 
predator with many prey, rather than improve the score of a predator 
with a single prey. Specialisation allows improved evolutionary adaptation. 

As the total population of the level below increases, then it becomes 
possible for predators with fewer prey to exist and still maintain 
their populations above the critical value. If highly specialised predators 
can exist then they will replace generalists because they are better 
adapted. This is why the transition from niche limited to population 
limited regimes occurs. If this argument is valid then species should 
become better adapted predators as trophic level decreases and $R$ 
increases. We investigate this by defining the mean score of species $i$ by
\begin{equation}
\bar S_{i} = \sum_{j = 0}^{S} f_{ij}S_{ij}.
\label{meanscore}
\end{equation}
This quantity is shown averaged over all the species on each trophic 
level for different $R$ values in \fig{r4levelfig}C. As expected, the average 
score on a trophic level mirrors the changes in predator specificity 
(\fig{r4levelfig}D). 


\begin{figure}[t]
\begin{center}
\rotatebox{0}{\scalebox{2.5}{\includegraphics[width=.3\textwidth]{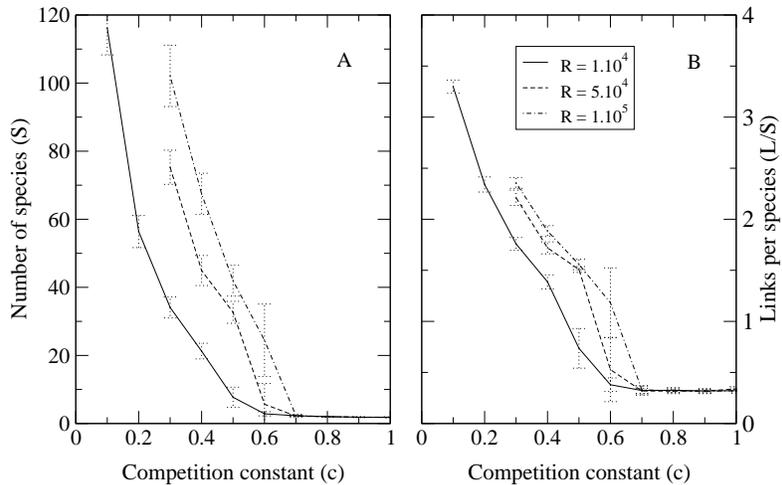}}}
\end{center}
\caption[]{The effect of the competition constant on species number (A) 
and links per species $L/S$ (B) for three different values of $R$. The 
details of these simulations are contained in the text. The error bars 
show standard deviations.} 
\label{c2fig}
\end{figure}


\subsection{Change in food-web structure with $c$}

We will now consider the effect of altering the competition constant 
$c$, which controls the strength of direct inter-specific competition in 
the model by parameterising the equation for the degree of 
interference competition, $\alpha_{ij}$, between two species that share 
the same resource (see Eq.~(\ref{c})). Increasing $c$ at a given feature 
overlap, $q_{ij}$, increases $\alpha_{ij}$, corresponding to greater 
competition. Given the importance of competition in the above arguments 
explaining the trophic level structure of the food webs, we might expect 
that $c$ should have a significant effect on the food web structure. We 
shall see that this is indeed the case.

The data set for this part of our study was obtained by performing 
twenty independent runs of the model for $c = 0.1,0.2,\ldots,0.4$ and ten 
runs for $c = 0.5, 0.6,\ldots,1.0$ with $R = 1\times 10^{4}$. In addition 
ten runs were performed at eight values of $c$: $c = 0.3, 0.4,\ldots,1.0$, 
for both $R = 5 \times 10^{4}$ and $R = 1 \times 10^{5}$. The other 
parameters were kept constant with $b = 0.03$. The simulations were 
run for 200000 speciation events, by which time the community properties 
had become time independent. The statistics presented here were obtained 
by the usual procedure of time averaging over the last 20000 iterations of 
each run and then ensemble averaging over the ten runs at each set of 
parameter values.

In \fig{c2fig} the species diversity $S$ and links per species $L/S$ are 
shown as a function of $c$ and $R$. The links per species is simply the 
number of predator-prey interactions in the food web ($L$) divided by the 
total species number ($S$). In calculating $L$ only links which 
constituted greater than 1\% of a predator's diet were counted. Both 
statistics increase rapidly once $c$ is smaller than some critical value. 
This indicates that there is some threshold competition level above 
which complex food webs cannot be evolved, but below which complexity in 
terms of species number and interactions increases rapidly. This critical 
value is around $c = 0.7$, and does not seem to depend strongly on $R$, 
although with just three different $R$ values spanning one order of 
magnitude it is difficult to be certain. 

The increased species number and links per species resulting from decreased 
interference competition arises from a number of intertwined processes.
Smaller $c$ values allow a greater number of species to exploit the same 
prey, which increases $L/S$ and also $S$, because each prey species can 
now support more predators which themselves provide food for other species. In 
addition, decreasing $c$ should decrease the size of the denominator in 
(\ref{g_ij}), resulting in increased feeding rates and allowing more species 
to exist with equilibrium populations above $N^{\rm min}$. 

The reason for the critical value in $c$ seems to be that above this 
point the probability of more than one species being able to exploit the 
same prey is very small. Thus the food webs are restricted to food chains, 
or at most one or two species on each trophic level. This has been checked 
by visual inspection of webs evolved at high $c$ values. The length of 
the food chains is determined by $\lambda$ and $R$, and is for the values 
used here at most four. Thus the food chains possess few species and 
few trophic links. 


\begin{figure}[t]
\begin{center}
\rotatebox{0}{\scalebox{2.5}{\includegraphics[width=.3\textwidth]{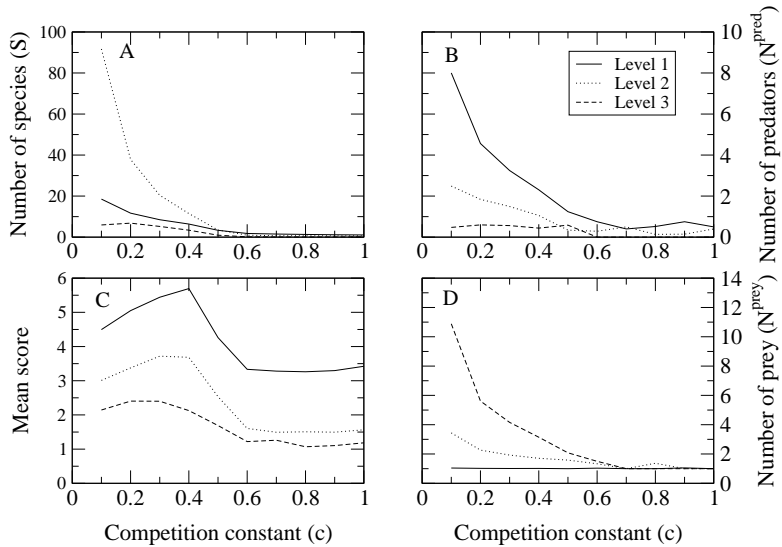}}}
\end{center}
\caption[]{The effect of the competition constant, $c$, on four quantities 
as a function of trophic level. A: the number of species on each trophic 
level. B: the number of predators of each species averaged over species 
on the same trophic level. C: the mean score averaged similarly. D: the 
number of prey again averaged over trophic level. These statistics are 
from the simulations with $R = 1\times10^4$ performed to investigate 
the effect of altering $c$ described in the text.} 
\label{cLevel}
\end{figure}


In explaining the trophic level structure of the model food webs and the 
effect of $R$ on that structure, we proposed that each trophic level can be 
placed on a continuum between prey population limited and niche limited 
regimes. We will now further explore these ideas by investigating the 
effect of $c$ on the trophic level structure of the food webs. In \fig{cLevel} 
we present four statistics averaged over all the species on the same 
trophic level as a function of $c$. It was mentioned above that reducing 
$c$ increases the number of species that can exploit the same prey. This 
is partially confirmed by \fig{cLevel}B, which shows average predator number 
increasing with decreasing $c$ values for the lower two levels, 
but not the top level. The latter however is exploited by only a few 
omnivorous predators. From \fig{cLevel}B we see that the average number 
of prey species per predator increases for trophic levels 2 and 3, but not
for trophic level 1, where species are specialised to feed off the environment 
for all $c$ values. This high specificity suggests that trophic level 1 
is niche limited. The average prey numbers for levels 2 and 3 imply that 
the generalist species on level 3 are prey population limited, whereas 
those on level 2 lie somewhere between the two extreme regimes. 

These observations explain the changes in trophic level occupation with 
decreased $c$ shown in \fig{cLevel}A. The reduced level of 
inter-specific competition allows more species to be specialised to feed 
off the environment so the occupation of level 1 increases. These provide 
more potential prey for the level 2 species and each level 1 species 
can support more predators. These two effects will compound each other, 
explaining the rapid increase in the occupation of trophic level 2. In
contrast, the number of level 3 species remains roughly constant, as the 
diversity of this level is limited by the total population of, rather than 
the number of niches in, the level below. 

The bottom left graph of \fig{cLevel} shows the mean score as defined by 
\eqr{meanscore} for each of the trophic levels as a function of $c$. This 
quantity has a hump shaped appearance for all three trophic levels. The 
increase in mean score for small $c$ values is probably because 
increasing $c$ reduces the disadvantage of high feature overlap values 
between groups of competing species, see Eq.~(\ref{c}). This allows groups 
of competitors to evolve towards possessing the same set of optimum features
for exploiting their mutual prey. The decrease in mean score as $c$ is 
further increased beyond the critical value is less explicable. It is 
probably related to the collapse in community size. 

\subsection{Change in food-web structure with $b$}

The last parameter we will consider is the saturation constant $b$, which
controls the strength of predator-prey interactions within the webs. 
This can be shown by rearranging the general form of 
the functional response (\ref{g_ij}) such that it is parameterised by 
$S_{ij}/b$. Thus $b$ scales the interaction scores. The effect of 
altering $b$ was determined by running ten independent simulations of 
the model at each of eleven values of $b$: 
$b = 0.00, 0.01,\ldots,0.10$. This was repeated for 7 different values of 
$R$: $R=10^{4},10^{4.25},\ldots,10^{5.5}$. In addition for $R=10^{5.75}$ 
ten simulations were performed at $b = 0.02,0.03,\ldots,0.10$ and for
$R=10^{6}$ the values $b=0.04,0.05,\ldots,0.10$ were used. For $R=10^{5.75}$ 
and $R=10^{6}$ it was not possible to implement the full range of $b$ values, 
as for small $b$ and large $R$ the webs become prohibitively large. This 
gives a total of 930 separate simulations. The other parameters were kept 
constant with $c = 0.5$. The simulations lasted for 120000 speciation 
events. The statistics were first time averaged over the last 20000 
iterations of each run, and then ensemble averaged over the ten runs at 
each set of parameter values.


\begin{figure}[t]
\begin{center}
\rotatebox{0}{\scalebox{2.5}{\includegraphics[width=.3\textwidth]{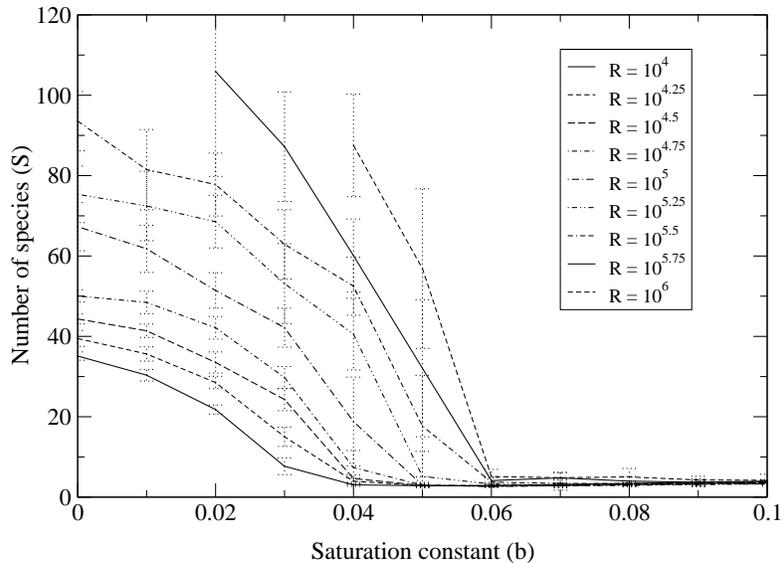}}}
\end{center}
\caption[]{The effect of the saturation constant, $b$, on species number $S$, 
for different values of $R$. The statistics are ensemble averages over 
ten runs at each set of parameter values. The error bars show the standard 
deviations for these ensembles. The details of the simulations and 
averaging procedure are contained in the text.} 
\label{bsfig}
\end{figure}


In \fig{bsfig} the number of species is shown for all parameter values. We 
can see from this that for any particular rate of resource input, $R$, 
the community size decreases with increasing $b$, until at some critical 
value of $b$ the webs collapse to just a few species. The webs remain 
small as $b$ is increased beyond this point. Increasing $R$ and keeping 
$b$ constant leads to increased size, as was seen in \fig{rsfig}. However 
the critical value of $b$ above which large webs cannot be assembled is 
only weakly dependent on $R$. It increases from around $b = 0.04$ to 
$b = 0.06$, as $R$ increases by two orders of magnitude.

The reason for the critical value is not clear. Certainly larger $b$ values 
will reduce the magnitude of the predator functional responses, perhaps 
making it difficult for a diverse collection of species to exist with 
equilibrium populations above $N^{\rm min}$. However it is hard to see 
why this should cause such sudden declines in species number with $b$, 
especially as some species do exist above the critical value. 

\medskip

In this section we have shown that the parameters $R$, $b$ and $c$ each 
have significant effects on the model food web structures. The parameter 
$R$ scales the food web size and impacts on the trophic level structure of the 
food webs. For the parameters $b$ and $c$ critical values exist above 
which large communities cannot be evolved. These critical values will 
depend on the other model parameters in ways not fully investigated here. 
However it was shown that they are relatively insensitive to $R$. In 
addition, we proposed a framework of niche limited and prey population 
limited regimes that could explain the changes in trophic level 
occupation with $c$ and $R$. In the next section we discuss the relevance 
of these results to real food webs.

\section{Comparison to empirical food webs}
\label{comparison}

The food webs constructed from simulations of the model have been 
compared to empirical food webs in earlier publications (Caldarelli 
et al., 1998; Drossel et al., 2001). The emphasis in these 
earlier papers was on comparing the percentage of species which were basal
(having no prey), intermediate (having both prey and predators) and top
(having no predators) and the percentage of predator-prey links which 
connected species of this type. Agreement was generally good, with the 
exception of the number of links per species which tended to be lower 
in model webs than in real ones. However, there are considerable problems 
with the direct comparison of model and empirical webs in this way, and
it is not even clear that comparisons of this type are really meaningful.
One problem is that real webs differ considerably in structure. It is 
not clear if this is because of the different nature of the community 
(marine, desert,...), differences in methodology, inadequate sampling, 
or a whole host of other reasons. The amount of time and effort required 
to get reliable data is formidable, and so the accuracy of much of the data 
is unknown. Some of these questions are discussed in a recent review of food
webs (Drossel and McKane, 2003; see also references therein). Another problem 
rests with the interpretation of the modelling process. In the current 
state of development of the model, it is unclear on what scale the 
predictions are expected to hold. On the one hand the model communities are 
quite small suggesting a local or regional scale, but on the other there is 
no immigration, a situation which may best correspond to a continental scale 
where diversity, for instance, is controlled by speciation and extinction 
(Rosenzweig, 1995). However, bearing these caveats in mind, it is 
nevertheless interesting to make a tentative comparison between the 
predictions of the model and empirical webs.

The three parameters $R$, $c$ and $b$ each have significant effects on 
model food web structure but of these three parameters only $R$ can be 
identified with any property of real ecosystems: it being reasonable to 
associate $R$ with the rate of input of limiting external resources such 
as light or nutrients. If this interpretation of $R$ is correct, then we 
would expect it to correlate with primary productivity, and in fact the 
relationship is almost exactly linear. Therefore we can compare the 
trends observed when altering $R$ in the model food webs with the trends 
observed in empirical food webs that span a range of primary productivities. 
There are several studies of the effect of productivity on species 
diversity and these can be compared with Figures \ref{rsfig} and 
\ref{rsplevelfig}A, which show the effect of $R$ on the species diversity 
of the complete food web and the individual trophic levels respectively. 

The species diversity of the individual trophic levels in the 
model are monotonically increasing saturating functions of $R$ 
(Fig.~\ref{rsplevelfig}A). The degree of saturation increases as the 
trophic level decreases so that level 1 or producer diversity is 
practically independent of $R$ for all but the smallest values. For animal
studies of regional species, diversity within a trophic group generally 
show unimodal i.e. hump-shaped dependence on productivity, so that diversity 
peaks at an intermediate productivity level (Mittelbach et al., 2001). 
Thus the model obtains the correct behaviour for low productivities, 
which is probably because the model incorporates the same energetic 
considerations as are used to construct the most widely accepted 
explanation for the increasing phase in the real diversity: the 
``species-energy'' hypothesis (Wright, 1983), but fails to predict the 
decrease in diversity at high productivities. This may indicate that the 
model lacks some important components but it could equally well be due 
to equating trophic levels with trophic groups. In addition, some studies have 
concluded that on a continental scale a monotonic relationship between 
diversity and productivity may be the dominant pattern (Waide et al., 
1999; Chase and Leibold, 2002).

For plants the empirical studies paint a similar picture with hump shaped 
patterns predominating on regional scales but with some evidence for 
monotonic relationships on larger scales (Waide et al., 1999; 
Mittelbach et al., 2001; Chase and Leibold, 2002). Neither possibility 
corresponds to the effectively constant number of level 1 species observed 
in Fig.~\ref{rsplevelfig}A. There are no empirical studies, at least known 
to the authors, comparable to Fig.~\ref{rsfig} where the total diversity 
is considered. Such a study might give the model behaviour, a monotonic 
relationship between total diversity and productivity. This remains to be seen.


\begin{table}
\centerline{
\begin{tabular}{cccccccc} \hline
& &  & \multicolumn{4}{c}{\it Trophic Level} & \\ 
{\it Ecosystem}& {\it Number of} & {\it Links per}  & {\it 1} & {\it 2} & 
{\it 3} & {\it 4} &{\it Reference}\\
\cline{4-7} 
{\it Name}&{\it species}  &{\it species}  & \multicolumn{4}{c}
{\it Number of species} & \\
& & & \multicolumn{4}{c}{\it Mean number of prey} & \\ 
\hline \hline
Bridge Brook Lake\dag& 75 & 7.37&39&34&2&---& Havens, 1992\\
& & &1.0&15.2&18.0&---&\\
\hline
Scotch Broom* & 154 & 2.40&1&24&117&12& Memmott {\it et al.}, 2000\\
& & &1.0&2.5&2.5&1.8&\\
\hline
Canton Creek\dag& 108 & 6.56&56&52&---&---& Townsend {\it et al.}, 1998\\
& & &1.0&13.6&---&---&\\
\hline
Chesapeake Bay*\dag & 33 & 2.18&5&15&13&---& Baird \& \\
& & &1.0&2.5&2.6&---&Ulanowicz, 1989\\
\hline
Coachella Valley*\dag & 30 & 9.67&3&22&5&---&Polis, 1991\\
& & &1.0&10.7&10.8&---&\\
\hline
El Verde Rainforest* & 156 & 9.68&28&98&28&2& Waide \& \\
& & &1.0&13.5&6.6&1.0&Reagan, 1996\\
\hline
Grassland \dag & 75 & 1.51&8&15&52&---& Martinez {\it et al.}, 1999\\
& & &1.0&1.1&1.9&---&\\
\hline
Little Rock Lake*\dag & 181 & 13.12&63&80&38&---& Martinez, 1991\\
& & &1.0&17.9&24.8&---&\\
\hline
Skipwith Pond*\dag & 35 & 10.86&1&18&16&---&Warren, 1989\\
& & &1.0&1.8&21.7&---&\\
\hline
St. Marks Seagrass*\dag & 48 & 4.60&6&31&11&---&Christian \& \\
& & &1.0&5.0&5.9&---&Luczkovich, 1999\\
\hline
St. Martin Island* & 44 & 4.95&6&29&6&3&Goldwasser \& \\
& & &1.0&5.0&11.7&1.3&Roughgarden, 1993\\
\hline
Stony Stream & 112 & 7.43&63&46&3&---&Townsend {\it et al.}, 1998\\
& & &1.0&17.9&2.7&---&\\
\hline
Ythan Estuary 2* & 92 & 4.58&5&44&42&1&Hall \& \\
& & &1.0&3.1&6.7&4.0&Raffaelli, 1991\\
\hline
Ythan Estuary 1* & 134 & 4.46&5&44&81&4&Huxham {\it et al.}, 1996\\
& & &1.0&3.1&5.6&2.0&\\
\hline \hline
\end{tabular}}
\caption{The number of species, links per species, distribution of species 
between trophic levels and mean prey number as a function of trophic level 
for fourteen food webs. The details of this data set are summarised in 
Dunne {\it et al.} (2002b). The ecosystems with hump shaped level 
distributions are marked by an asterisk (*) and those with decreased 
predator specialisation with trophic level by a dagger (\dag).
\label{empirical}}
\end{table}


There are other predictions from the previous section, regarding model 
food web trophic structure, that are independent of the model parameters. 
We shall focus on two of these: the hump shaped distribution of species 
between the trophic levels and the increase in predator specialisation as 
trophic level decreases. These patterns were explained in the model ecosystems
using the ideas of niche limited and prey population limited regimes. If 
such ideas are applicable to real webs we would expect to find the same 
patterns in the empirical data. In \tabl{empirical} the number of species 
occupying each trophic level, calculated as the minimum path length from 
the species to the environment, and the number of prey averaged over all 
species on the same trophic level, counting the environment as a prey 
species, are shown together with links per species and total number of 
species of the whole food web for fourteen of the largest and most highly 
resolved food webs in the current literature. The statistics were
calculated using taxonomic rather than trophic species since although 
using the former may reduce methodological bias (Briand and Cohen, 1984), 
the latter are the functional units in the food webs. These webs span a 
range of sizes and habitats, the details of which are contained in the 
individual references or summarised in Dunne et al. (2002b).

Of these fourteen ecosystems, ten have a distribution of species between 
levels that peaks at an intermediate trophic level number, indicated by 
an asterisk (*) in \tabl{empirical}. This is more than might be expected 
by chance, assuming that any level is equally likely to be the modal one. 
However from this we cannot conclude that the trophic level structure of 
real food webs is controlled by an interplay between number of available 
niches on the lower levels and resource availability on the higher levels, 
as was proposed for the model food webs. This is because many other 
hypotheses are likely to be consistent with this data, for instance, it is 
easy to show that similar hump-shaped trophic level distributions are 
obtained if ecosystems are generated with links apportioned randomly.
However this result is reassuring, and this feeling is reinforced by the 
number of trophic levels possessed by the food webs in \tabl{empirical}, 
there being only one two level ecosystem, all others having either three 
or four levels. This is also consistent with model ecosystems where, 
except in the case of very low $R$ values, there are always at least 
three levels and for larger $R$ generally four. This can be seen from 
Fig.~\ref{rsplevelfig}A.

The status of the second proposed pattern, the increase in predator 
specialisation with decreased trophic level, is not so clear. It holds 
for eight of the fourteen webs, those marked with a dagger ($\dag$) 
in \tabl{empirical}, but for two of these eight webs, Chesapeake Bay 
and Coachella Valley, the difference between the mean prey number on 
the second and third levels is very small. Given that the mean prey 
number for the first trophic level is always going to be 1.0, guaranteeing 
that a two a level food web will display the pattern, we conclude that 
the current state of the data is such that a meaningful comparison is not
warranted.   

Finally, the number of links per species $L/S$, given in the third column 
of \tabl{empirical}. For the model webs this quantity depends on the 
parameters used. For the parameters $R = 1 \times 10 ^ 5$, $c = 0.5$, 
$b = 0.005$, the long time average number of links per species is 1.69. 
This value increases as $c$ decreases, as can be seen from 
Fig.~$\ref{c2fig}$B, but it rarely approaches the sorts of values observed 
in \tabl{empirical}. This confirms the earlier findings that the model 
food webs are link poor compared to real food webs. 

The problems we have encountered in this section when making comparisons
between model and empirical webs have been typical. A major difficulty is
that a property of model webs may be seen in a number of empirical webs, 
perhaps even a large majority of them, but it is not clear if those in
which the property is absent are different in some way, or if it is 
simply that the data is not good enough. For this reason, it may prove 
more productive to seek out properties of real food webs which are more 
universal, and so allow more meaningful comparisons with model webs. We
now move on to the investigation of a property which may be of this type:
the distribution of link strengths in food webs. 
 
\section{Interaction strengths in model food webs}
\label{interactions}

The majority of the food web statistics examined in the previous sections 
take no account of the strength of the predator-prey interactions
in the model generated communities. They describe properties of binary food 
webs for which a link is either present or absent. This approach was adopted 
for reasons of simplicity and to facilitate comparison with the empirical data,
where for most well resolved food webs the strengths of the interactions 
are not known. For those natural communities for which interaction strengths 
have been quantified, significant variation in link importance has been 
observed and a putative general pattern of ``a few strong interactions 
embedded in a majority of negligible effects'' proposed (Paine, 1992; Fagan 
and Hurd, 1994; Raffaelli and Hall, 1996; Wootton, 1997), although this 
emerging consensus has not remained unchallenged (Sala and Graham, 2002). It 
was soon realised that the variation of interaction strengths within ecosystems
could have important implications for their functioning. In particular, it 
has been shown that using interactions drawn from realistic distributions 
increases the probability that a random model food web will be stable 
(Yodzis, 1981). This idea that the observed pattern of interaction strengths 
could be important for community stability was refined by suggestions that 
the pattern of variation in interaction strengths between trophic levels 
(de Ruiter et al., 1995), the preponderance of weak interactions 
(McCann et al., 1998) and the presence of weak omnivorous 
interactions in long loops (Neutel et al., 2002) could all be important 
factors in determining ecosystem stability. In light of these ideas and 
observations we will now investigate the distribution of link strengths 
generated by the model.

We will begin by considering the distribution of the efforts, $f_{ij}$, in 
the model food webs. These quantities, defined by (\ref{ESS}), give the 
fraction of an individual predator's diet obtained from a particular prey. 
Thus they do not represent interaction strengths but rather diet 
contributions. Just because a predator consumes only one particular prey 
does not necessarily imply that the effect of the predator on that prey 
will be high. In fact it has been shown in real food webs that the
percentage of a particular prey in a predator's diet does not correlate 
strongly with interaction strength (Wootton, 1997). However since our 
criterion for including a link in the previous section was that $f_{ij}$ 
should be greater than 0.01, and because $f_{ij}$ is a quantity with an 
unambiguous meaning relevant to real food webs, then investigating the 
distribution of $f_{ij}$ is a worthwhile exercise. 


\begin{figure}[t]
\begin{center}
\rotatebox{0}{\scalebox{2.5}{\includegraphics[width=.3\textwidth]
{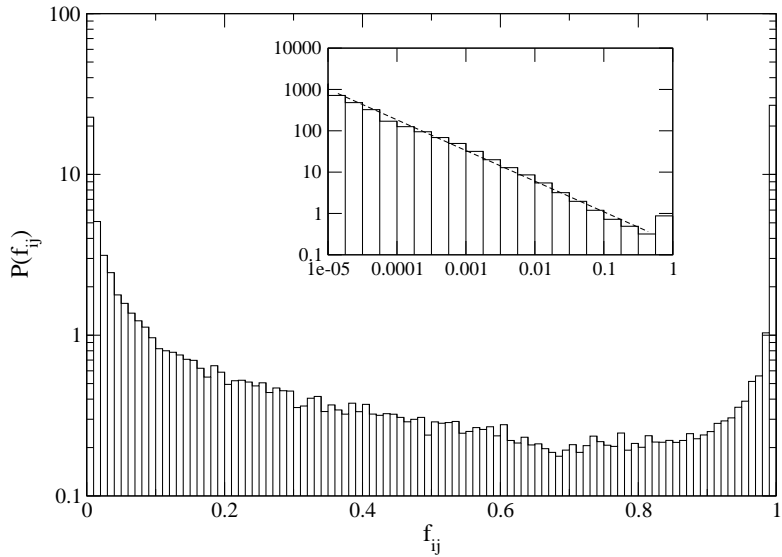}}}
\end{center}
\caption[]{The distribution of efforts or 
predator diet fractions $f_{ij}$ averaged over 400 model communities 
evolved for 120000 iterations with the parameters $R = 1\times 10^5$, 
$b = 0.005$ and $c = 0.5$. The inset shows the same data but with both 
axes logarithmically scaled and logarithmically spaced bins. 
The power law fit shown for the inset has an exponent of $-0.74 \pm 0.01$.} 
\label{fdist}
\end{figure}


In Fig.~\ref{fdist} the probability distribution of the efforts is shown 
averaged over the final communities from four hundred independent runs, 
with the same parameters, of the model. The probability distribution 
is calculated for all predator-prey links with $f_{ij} > 10^{-6}$. This 
minimum cut off was applied because this is the smallest value that $f_{ij}$ 
is allowed to take if $S_{ij} > 0$ (Drossel et al., 2001). Thus 
we have only included efforts whose values were determined by iterating 
(\ref{ESS}) and (\ref{g_ij}), rather than by the limitations of our algorithm.

From the main graph of Fig.~\ref{fdist} it is clear that $P(f_{ij})$ is 
heavily biased towards the limits of its possible range, $f_{ij} = 1$
and $f_{ij} = 0$. The probability that a randomly chosen link has an effort 
in between these two limits is low. This effect is actually more dramatic 
than it appears, since the vertical axis of this graph has been scaled 
logarithmically. There is also a slightly greater weighting towards small 
$f_{ij}$ values: 60\% of links have $f_{ij} \le 0.5$. The inset
graph of Fig.~\ref{fdist} shows $P(f_{ij})$ plotted with both axes scaled 
logarithmically and using bin intervals that are also logarithmically
spaced. This reveals that the probability distribution of $f_{ij}$, for 
small $f_{ij}$, is well described by a power law with an exponent of $-0.74$. 

It seems that predation patterns in the model are highly uneven 
with most species possessing one main prey, but including very small amounts 
of other species in their diets. The fraction of species that predate 
several species evenly, resulting in intermediate $f_{ij}$ values, must 
be low. A similar shaped distribution of efforts to this, albeit with a 
greater proportion of small values, has been observed in a food web model 
with adapting but not evolving foragers (Kondoh, 2003). This pattern can also 
be compared to a detailed study of a detritus based stream food web, where 
the dietary importance of links to a consumer were quantified with a 
measure similar to $f_{ij}$ (Tavares-Cromar and Williams, 1996). Links were 
characterised as weak, moderate, strong and very strong depending on the 
fraction of the predators diet represented by the link. The percentage of 
links in these categories were 42-52\%, 17-25\%, 5-13\% and 21-25\% 
respectively, indicating a significant bias towards weak links and some bias 
towards very strong links as well. However it is difficult to reconstruct 
the shape of the true distribution from just four categories whose 
boundaries are inevitably arbitrary. 

In Fig.~\ref{fdist} we have shown $P(f_{ij})$ at only one set of parameter 
values. However this pattern is remarkably insensitive to the choice
of parameters. The probability distribution of $f_{ij}$ was calculated for 
a large range of $b$, $c$ and $R$ values using the webs generated in the 
previous section, and the same bias to $f_{ij} = 0$ and $f_{ij} = 1$
was found in all instances where the communities were reasonably large. In 
particular, the power law distribution for small $f_{ij}$ was found with 
the same exponent, $0.725 \pm 0.025$, for all values of the rate of 
resource input, $R$, and the saturation constant, $b$, that produced food 
webs with more than a few species. The distribution was found to depend 
slightly on $c$. As the competition constant is reduced there is an 
increased probability of finding predator-prey links with intermediate $f_{ij}$
values. This is what we would expect from the findings of the previous 
section: decreased $c$ results in reduced competition, allowing predators 
to exploit a greater range of prey species.

As mentioned above, the effort $f_{ij}$ associated with a predator-prey 
interaction may not be a good measure of its strength. The interaction 
score $S_{ij}$ is a better candidate since it parameterises the predator 
functional response (\ref{g_ij}). However this quantity has the drawback 
of being relevant only to our particular choice of functional response. 
In addition, since interference competition is incorporated into (\ref{g_ij}),
competition is a direct interaction in the model the strengths of which it 
would be useful to quantify with the same measure as used for the 
predator-prey interactions. For these reasons we shall 
use as our definition of interaction strength, the Jacobian or 
``community matrix'' $J$ (May, 1973). This is an $S \times S$ matrix with 
elements $j_{ij}$ defined by:
\begin{equation}
j_{ij} = \left(\frac{\partial H_{i}(N_{1}(t),N_{2}(t),\ldots,N_{S}(t))}
{\partial N_{j}(t)}\right)^{*},
\end{equation}
where $H_{i} \equiv dN_{i}/dt$ is the net growth rate of 
species $i$ (\ref{balance}). This is a function of the population 
densities of all species that directly interact with $i$. The asterisk 
indicates that the partial derivatives are to be calculated at an 
equilibrium point. The latter is defined as a set of population densities 
that give zero growth rates, i.e. 
$H_{i}(N^{*}_{1},N^{*}_{2},\ldots,N^{*}_{S}) = 0$ for all $i$. This 
restriction to an equilibrium point means that the elements of $J$ are the 
coefficients of the linearization of the population dynamics about this 
point, and the perturbations from equilibrium, 
$x_{i}(t) = N_{i}(t) - N_{i}^*$, obey
\begin{equation}
\frac{dx_{i}(t)}{dt} = \sum_{j = 0}^{S} j_{ij} x_{j}(t),
\end{equation} 
to first order. Therefore the community matrix determines the local stability 
of the equilibrium point. If all its eigenvalues have negative real parts, the 
system will return asymptotically to the equilibrium after any infinitesimally 
small perturbation of the population densities.

The method for calculating the Jacobian matrix of a model food web 
involves two steps. Firstly, location of an equilibrium point: this was 
simple as almost always the population dynamics integrated to a 
stable equilibrium suitable for the calculation of the Jacobian. On the 
rare occasions when such an equilibrium was not found, reducing the size 
of the integration time step resolved the problem. Secondly, calculation 
of the partial derivatives: this was more complicated as it is not possible to
derive analytic expressions for the partial derivatives by differentiating 
(\ref{g_ij}). This is because the $f_{ij}$, which we do not have an explicit 
expression for, will also depend on the population densities. This problem 
was circumvented by numerically calculating the partial derivatives using 
Ridders' method (Ridders, 1982; Press et al., 1988). This involves 
calculating the change in species growth rates resulting from increasingly 
small perturbations of one of the population densities, the other densities 
remaining unchanged, and then extrapolating from these values to estimate 
the result of a infinitesimal perturbation corresponding to the partial 
derivative. During the calculation of the adjusted growth rates the efforts 
were allowed to change from their values at the population equilibrium 
such as to always satisfy (\ref{ESS}), thus generating Jacobian elements 
for species that are adaptively changing their foraging efforts.


\begin{figure}[t]
\begin{center}
\rotatebox{0}{\scalebox{2.5}{\includegraphics[width=.3\textwidth]{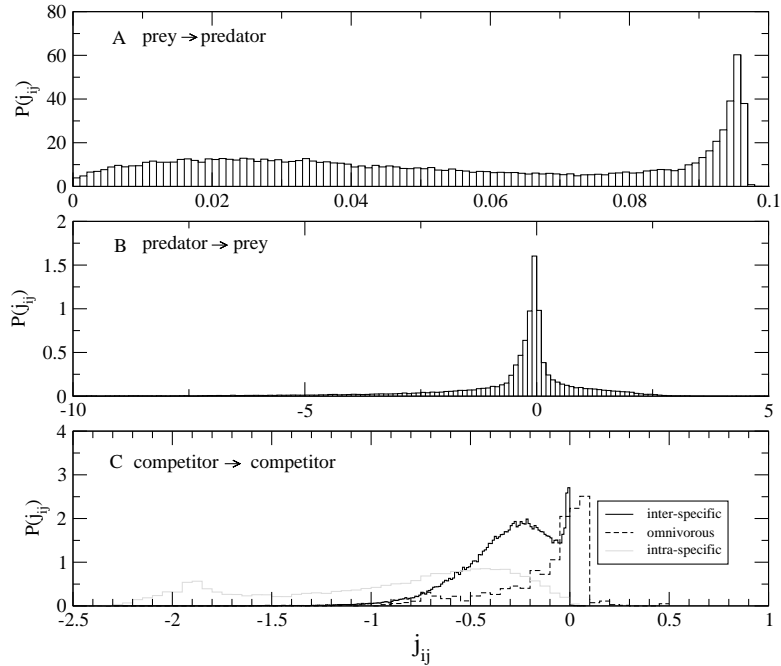}}}
\end{center}
\caption[The distribution of Jacobian matrix elements]{The distribution of the elements of the 
Jacobian matrix $j_{ij}$ for the effect of a prey 
species on its predator (A), the effect of a predator on its prey (B) and 
the effect of competitors on each other (C). The latter is sub-divided into 
inter-specific, omnivorous and intra-specific interactions. These results 
were obtained by averaging over 400 communities evolved for 120000 iterations 
with the parameters $R = 1\times 10^5$, $b = 0.005$ and $c = 0.5$.} 
\label{Jdist}
\end{figure}


The Jacobian matrices were calculated numerically for the four hundred 
food webs used to generate Fig.~\ref{fdist}. The distribution of all non-zero
$j_{ij}$ values was found to be strongly skewed towards weak interactions, 
which is perhaps unsurprising given the distribution of $f_{ij}$ values 
discussed above. In order to better understand the structure of the Jacobian 
matrices we then examined the distribution of $j_{ij}$ values associated 
with specific direct interactions. Firstly, we examined the effect of a 
prey species on a predator for all predator-prey interactions where the 
prey formed greater than 1\% of the predator's diet and the predator and 
prey were not also competitors. This is shown in Fig.~\ref{Jdist}A. It 
shows a broad distribution of values between 0 and 0.1, with a peak lying 
just below the upper limit of this range. That the lower limit of this 
distribution is zero derives from the fact that the direct effect of a 
prey on a predator is always positive. The upper limit can be understood 
as follows. We would expect that the impact of a prey on a predator will 
be greatest when the prey is the only species in the predator's diet. In 
this case indexing the predator as $i$ and the prey as $j$, we have using 
(\ref{1p_1p})
\begin{equation}
j_{ij} = \lambda N^{*}_{i}\left(\frac{\partial g_{ij}(t)}
{\partial N_{j}(t)} \right)^*= 
\lambda N_{i}^*\frac{S_{ij}^2 N_{i}^*}{(bN_{j}^* + S_{ij}N_{i}^*)^2}.
\label{jij}
\end{equation}
For a given predator population density 
this will be maximised in the limit of low effective equilibrium prey 
population density in the ratio-dependent functional response, 
$bN_{j}^* \ll S_{ij}N_{i}^*$, which gives $j_{ij} = \lambda$ and 
$\lambda = 0.1$ for this simulation. The peak indicates that a significant 
number of predator-prey interactions are close to this prey limited regime. 
In Fig.~\ref{Jdist}B the corresponding distribution of Jacobian matrix 
elements for the effect of predators on prey is shown. This distribution is 
peaked around zero, but with a significant probability of values with a 
magnitude large compared to those for the effect of prey on predators. 
These values have both positive and negative sign. This last fact is 
quite striking and indicates that in many instances the direct effect of a 
predator on its prey is positive. It arises from the interference terms 
in (\ref{g_ij}). This means that mutualistic interactions, at least at 
equilibrium, occur within the model with $j_{ij} > 0$ and 
$j_{ji} > 0$. For the four hundred food webs considered here roughly 
one third of all predator-prey interactions are mutualistic. 

The bottom graph in \fig{fdist} (labelled C) gives the distribution of 
Jacobian matrix elements for interactions between competitors. These have 
been subdivided into three types denoted inter-specific, omnivorous and 
intra-specific. The first category refers to all competitive interactions 
between different species that are not also predator-prey pairs. A competitive
interaction is considered to occur if the two predators both obtain more 
than 1\% of their diet from the same prey species. These interactions are 
almost always negative. It is not apparent from the graph, but very 
occasionally a species has a positive effect on its competitor, probably 
as a result of adaptive foraging; they have a broad distribution that has 
local maxima at both zero and an intermediate interaction strength value. 
The second category, omnivorous, refers to all competitive interactions 
between different species that are also predator-prey pairs. This is a 
slight misnomer, as actually this definition encompasses only a subset of 
the omnivorous interactions in the webs. They were considered separately 
in order not to confuse the results for strictly competitive interactions. 
The last category, intra-specific, refers to interactions within a species, 
corresponding to the diagonal elements of the Jacobian matrix. These values 
are also always negative but on average have a larger magnitude than the 
inter-specific competitive interactions. These negative diagonal elements 
play an important role in generating locally stable equilibrium points 
(May, 1973).


\begin{figure}[t]
\begin{center}
\rotatebox{0}{\scalebox{2.5}{\includegraphics[width=.3\textwidth]{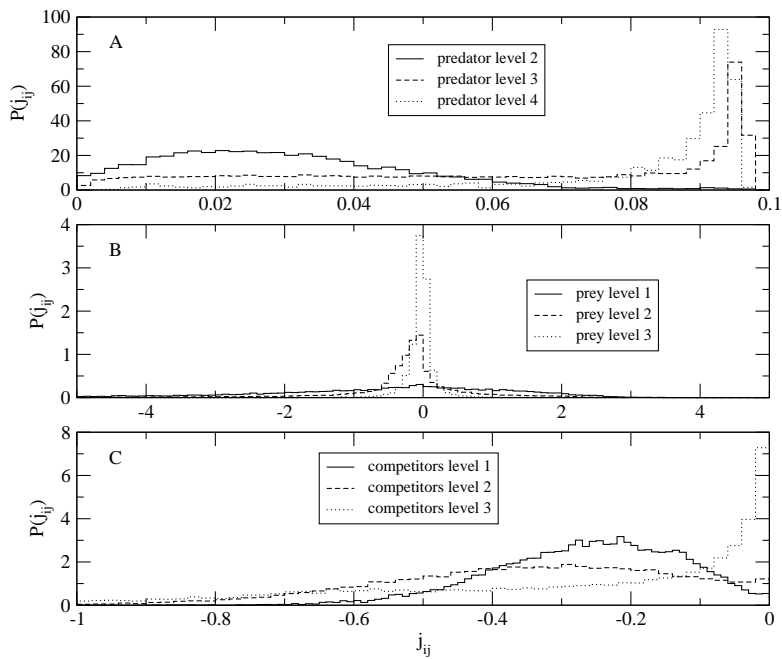}}}
\end{center}
\caption[The distribution of Jacobian matrix elements subdivided by trophic 
level]{The distribution of the elements of the Jacobian matrix $j_{ij}$ 
for the effect of a prey species on its predator (A), the effect of a 
predator on its prey (B), the effect of one competitor on another (C). 
The results are subdivided by the trophic level of the predator and prey 
respectively. These results were obtained by averaging over 400 communities 
evolved for 120000 iterations with the parameters $R = 1\times 10^5$, 
$b = 0.005$ and $c = 0.5$.} 
\label{JLevel}
\end{figure}


In Fig.~\ref{JLevel}A we show the Jacobian matrix elements for the 
effects of a prey on a predator, as in Fig.~\ref{Jdist}A, but now with the 
interactions subdivided by the trophic level of the predator species. 
Similarly Fig.~\ref{JLevel}B shows the distribution of Jacobian matrix 
elements for the effects of predators on prey subdivided by the prey 
trophic level, and Fig.~\ref{JLevel}C the elements between competitors on 
the same level. The results are quite striking. As predator trophic level 
increases, the distribution of $j_{ij}$ values shift towards the upper 
limit of $\lambda = 0.1$, and the effect of prey species on their predators 
increases. Conversely as prey trophic level increases the distribution 
of the effect of predator on prey becomes more strongly peaked at zero, 
corresponding to a decrease in the average magnitude of the interaction 
strength. The situation is less clear-cut for competitive interactions,
which become somewhat stronger from levels 1 to 2, and then show an increase 
in weak interactions for level 3. These changes in interaction strength 
must derive from the decrease in population size with trophic level shown in 
Fig.~\ref{rsplevelfig}B, but understanding exactly how is difficult. 
Semi-quantitatively, for the effects of prey on predators, we see that 
the right-hand side of (\ref{jij}) will be become larger until it saturates 
at a value $\lambda$, as the prey population becomes small.

These results indicate that species populations are controlled by predators 
and competitive interactions on the lower trophic levels, but are limited 
by prey population size on the upper levels. This fits with the theory 
of niche limited and prey population limited regimes advanced in Section 
\ref{structure}. The same patterns of increasing effects of prey on 
predators and decreasing impacts of predators on prey with trophic level 
has been found in real food webs (de Ruiter et al., 1995). They 
propose that this pattern stabilises food webs structures by ensuring that 
long loops must contain weak links hence reducing their negative impact 
on local stability (Neutel et al., 2002). It is possible that the 
same effect operates here to stabilise the model communities.


\begin{figure}[t]
\begin{center}
\rotatebox{0}{\scalebox{2.5}{\includegraphics[width=.3\textwidth]{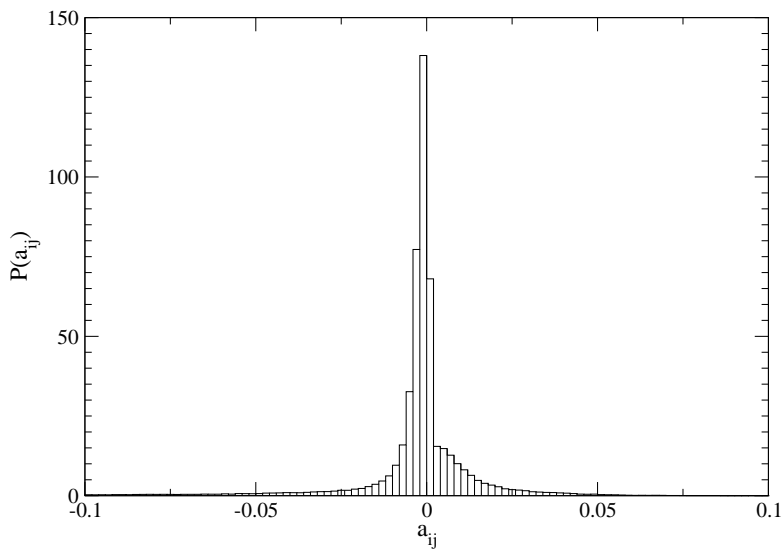}}}
\end{center}
\caption[The distribution of per-capita interaction strengths]{The distribution of the 
elements of the per-capita interaction matrix $A$ 
averaged over 400 model communities evolved for 120000 iterations with the 
parameters $R = 1\times 10^5$, $b = 0.005$ and $c = 0.5$.}
\label{Adist}
\end{figure}


To conclude this section we will discuss whether the distribution of 
interaction strengths observed in the model food webs fit
the proposed empirical pattern of many weak and few strong interactions 
referred to above. The elements of the Jacobian matrix, $j_{ij}$, measure 
the direct effect of an individual of species $j$ on the total population 
of species $i$ (Laska and Wootton, 1998). In empirical studies, per-capita 
interaction strengths are more often quoted. These are readily obtained 
in our case by defining the per-capita interaction strength matrix $A$ 
with elements $a_{ij} = j_{ij}/N^{*}_{i}$. The distribution of elements 
in this matrix averaged over the four hundred communities considered 
throughout this section is shown in Fig.~\ref{Adist}. Only those elements 
associated with a predator-prey interaction where the prey forms greater 
than 1\% of the predator's diet, or a competitive interaction where the shared
prey comprises greater than 1\% of both competitors diet, were used to 
calculate this distribution. This was done to avoid any implicit bias 
towards weak interactions. This distribution is clearly heavily skewed 
towards zero and thus fits the most commonly observed empirical pattern.

\section{Discussion}
\label{discussion}

In this paper we have reported the results of performing a very large 
number of simulations on a model of a coevolving multispecies community in
order to generate food webs. The results obtained were considerably more 
extensive than those reported in earlier publications. They were also 
complementary: in Drossel et al. (2001) the emphasis was on comparing 
model webs with empirical webs by looking at the number of top, intermediate 
and basal species and the links between them. The aim was to find values of 
$R, c$ and $b$ which gave agreement with particular empirical webs. It was 
found that, while values of these parameters could indeed be found which 
gave good agreement between the model and particular empirical webs, the data 
was not sufficiently consistent to warrant such detailed fitting. The aim 
in this study has had a different focus: to see how the model webs change 
their nature as the parameters $R, c$ and $b$ are varied. We will not repeat 
the conclusions of Section 4, except to draw attention to critical values 
of $c$ and $b$ above which the growth of complex webs does not seem possible. 
Another change in focus was the classification of species in terms of the 
level they occupied, rather than whether they were basal, intermediate or 
top species. Several trends were identified, but once again comparisons 
with data was problematical. We also listed other quantities (see Table 1) 
as parameters of the model, but which we did not vary. Of course, exactly 
what constitute ``parameters of the model'' is not well-defined, but in 
any case it would be interesting to carry out an investigation, not as 
detailed as the one reported here, in which $N^{\rm min}$ and 
$N^{\rm child}$ are changed from their present value of 1. We hope to 
carry out such a study in the near future.

In order to facilitate the comparison of model and empirical webs it would 
be useful to identify a property, trend or attribute which was shared by 
all empirical webs. One possibility is the existence of many weak links, 
with perhaps the distribution of weak link strengths resembling a power 
law. In Section 6 we showed that such a distribution was present in our model. 
This is a highly nontrivial result, since these distributions are an emergent 
property of the system, and not put in by hand as in some other work. It is 
a strong indication that weak links are the natural outcome of long-term 
community evolution coupled to population dynamics. The main difficulty here
lies with the definition of link strengths, of which there are a large number, 
both theoretically and empirically (Berlow et al., 2004). Fortunately, 
there are indications that the ``many weak links'' result holds independently 
of the precise definition of link strength.

The collection of empirical food web data is an extremely difficult, time 
consuming and labour intensive task. The data is improving, but it still 
remains difficult to know whether differences in structure between food 
webs are due to real effects, differences in methodology or inadequate 
sampling. We have mentioned one way forward --- the identification of novel 
effects which seem to be present in all webs. Another approach is to 
develop the model further by including additional structure which may give 
rise to different kinds of webs which mirror those found empirically. A 
prime candidate is the degree of isolation of a community. We have already 
commented on possible differences between the structures of communities on 
a continental scale and more local ones in Section 5. We plan to explore 
this aspect in more detail in the future.

In summary, the method of constructing food webs dynamically from the 
model studied in this paper seems to give food webs which are similar to 
empirical webs. This is particularly true for link strength distributions, 
but less so for the number of links per species. The difficulty in extracting 
universal attributes from food web data suggests that a way forward may be 
the further development of the model so as to produce a greater diversity 
of food webs.

\vspace{1cm}

\noindent{\bf Acknowledgements}: We wish to thank Jennifer Dunne, Mark 
Huxham and Phillip Warren for supplying food web data. CQ wishes to thank 
the EPSRC (UK) for financial support during the early part of this work.

\newpage

\section*{References}

\noindent Albert, R., Barab\'asi, A-L., 2002. Statistical mechanics
of complex networks. Rev. Mod. Phys. 74, 47-97.

\noindent Arditi, R., Michalski, J., 1996. Nonlinear food web
models and their responses to increased basal productivity. In:
Polis, G.~A., Winemiller, K.~O. (Eds.), Food webs: Integration of patterns and dynamics. 
Chapman and Hall, New York, pp.\,122-133.

\noindent Baird, D., Ulanowicz, R.~E., 1989. The seasonal dynamics of 
the Chesapeake bay ecosystem. Ecol. Monogr. 59, 329-364.

\noindent Bastolla, U., L\"{a}ssig, M., Manrubia, S.~C., Valleriani, A., 
2002. Dynamics and topology of species networks. In:  L\"assig, M., Valleriani, A. (Eds.),
Biological Evolution and Statistical Physics. Springer-Verlag, Berlin, pp.\,299-311.

\noindent Berlow, E.~L., Neutel, A-M., Cohen, J.~E., de Ruiter, P.~C., Ebeman, B., Emmerson,
 M., Fox, J.~W., Jansen, V.~A.~A., Jones, J.~I., Jonsson, T., Kokkoris, G.~D., Logofet, D.~O.,
 McKane, A.~J., Montoya, J.~M., Petchey, O., Raffaelli, D.~G., 2004.
 Interaction strengths in food webs: issues and opportunities. J. Anim. Ecol. in press.

\noindent Briand, F., Cohen, J.~E., 1984. Community food webs have 
scale-invariant structure. Nature 307, 264-267.

\noindent Caldarelli, G., Higgs, P.~G., McKane, A.~J., 1998. Modelling 
coevolution in multispecies communities. J. Theor. Biol. 193, 
345-358.

\noindent Chase, J.~M., Leibold, M.~A., 2002. Spatial scale dictates 
the productivity-biodiversity relationship. Nature 416, 427-430.

\noindent Christian, R.~R., Luczkovich, J.~J., 1999. Organizing and 
understanding a winter's seagrass foodweb network through effective trophic 
levels. Ecol. Model. 117, 99-124. 

\noindent Cohen, J.~E., Briand, F., Newman, C.~M., 1990. Community 
food webs, Biomathematics Vol. 20. Springer-Verlag, Berlin.

\noindent de Ruiter, P.~C., Neutel, A-M., Moore, J.~C., 1995. Energetics,
 patterns of interaction strengths, and stability in real ecosystems. 
Science 269, 1257-1260.

\noindent Drossel, B., Higgs, P.~G., McKane, A.~J., 2001. The influence 
of predator-prey population dynamics on the long-term evolution of 
food web structure. J. Theor. Biol. 208, 91-107.

\noindent Drossel, B., McKane, A.~J., 2003. Modelling Food Webs. In:
 Bornholdt, S., Schuster, H.~G. (Eds.), Handbook of graphs and networks.
 Wiley-VCH, Berlin, pp. \,218-247.

\noindent Dunne, J.~A., Williams, R.~J., Martinez, N.~D., 2002a. Food-web
structure and network theory: The role of connectance and size. Proc.
Natl. Acad. Sci. USA 99, 12917-12922.

\noindent Dunne, J.~A., Williams, R.~J., Martinez, N.~D., 2002b. Network 
structure and biodiversity loss in food webs: robustness increases with 
connectance. Ecol. Lett. 5, 558-567.

\noindent Fagan, W.~F., Hurd, L.~E., 1994. Hatch density variation of 
a generalist arthropod predator: population consequences and community 
impact. Ecology 75, 2022-2032.

\noindent Goldwasser, L., Roughgarden, J., 1993. Construction and analysis 
of a large Caribbean food web. Ecology 74, 1216-1233.

\noindent Hall, S.~J., Raffaelli, D., 1991. Food-web patterns: lessons 
from a species-rich web. J. Anim. Ecol. 60, 823-842.

\noindent Havens, K., 1992. Scale and structure in natural food webs. 
Science 257, 1107-1109.

\noindent Huxham, M., Beaney, S., Raffaelli, D., 1996. Do parasites reduce 
the chances of triangulation in a real food web? Oikos 76, 284-300.

\noindent Kondoh, M., 2003. Foraging adaptation and the relationship 
between food web complexity and stability. Science 299, 1388-1391.

\noindent Laska, M.~S., Wootton, J.~T., 1998. Theoretical concepts and 
empirical approaches to measuring interaction strength. Ecology 79, 
461-476.

\noindent L\"assig, M., Bastolla, U., Manrubia, S.~C., Valleriani, A., 
2001. Shape of ecological networks. Phys. Rev. Lett. 86, 4418-4421.

\noindent Martinez, N.~D., 1991. Artifacts or attributes: effects of 
resolution on the Little Rock Lake food web. Ecol. Monogr. 61, 
367-392. 

\noindent Martinez, N.~D., Hawkins, B.~A., Dawah, H.~A., Feifarek, B.~P., 
1999. Effects of sampling effort on characterization of food-web 
structure. Ecology 80, 1044-1055. 

\noindent May, R.~M., 1973. Stability and complexity in model ecosystems. 
Princeton University Press, Princeton.

\noindent Maynard Smith, J., 1974. Models in ecology.
 Cambridge University Press, Cambridge.

\noindent McCann, K., Hastings, A., Huxel, G.~R., 1998. Weak trophic 
interactions and the balance of nature. Nature 395, 794-798. 

\noindent Memmott, J., Martinez, N.~D., Cohen, J.~E., 2000. Predators, 
parasitoids and pathogens: species richness, trophic generality and body 
sizes in a natural food web. J. Anim. Ecol. 69, 1-15.

\noindent Mittelbach, G.~G., Steiner, C.~F., Scheiner, S.~M., Gross, K.~L., Reynolds, H.~L., Waide, R.~B., Willig, M.~R.,
 Dodson, S.~I., Gough, L., 2001. What is the observed 
relationship between species richness and productivity? Ecology 82, 2381-2396. 

\noindent Neutel, A-M., Heesterbeek, J.~A.~P., de Ruiter, P.~C., 2002. 
Stability in real food webs: Weak links in long loops. Science 296, 
1120-1123.

\noindent Paine, R.~T., 1992. Food web analysis through field measurement 
of per-capita interaction strength. Nature 355, 73-75.

\noindent Polis, G.~A., 1991. Complex trophic interactions in deserts: 
an empirical critique of food-web theory. Am. Nat. 138, 123-155.

\noindent Press, W.~H., Teukolsky, S.~A., Vetterling, W.~T., Flannery, B.~P., 
1988. Numerical recipes in C: the art of scientific computing. Cambridge 
University Press, Cambridge, pp.\,186-189.

\noindent Quince, C., Higgs, P.~G., McKane, A.~J., 2002. Food web structure 
and the evolution of ecological communities. In:  L\"assig, M., Valleriani, A. (Eds.),
Biological Evolution and Statistical Physics. Springer-Verlag, Berlin, pp.\,281-298.

\noindent Raffaelli, D.~G., Hall, S.~J., 1996. Assessing the relative 
importance of trophic links in food webs. In: Polis, G.~A., Winemiller, K.~O. (Eds.),
 Food webs: Integration of patterns and dynamics. Chapman and Hall, New York, pp.\,185-191.

\noindent Ridders, C.~J.~F., 1982. Technical note: accurate computation 
of $F'\{x\}$ and $F'\{x\}F''\{x\}$. Adv. Eng. Softw. 4, 75-76.

\noindent Rosenzweig, M.~L., 1995. Species diversity in space and time. 
Cambridge University Press, Cambridge.

\noindent Roughgarden, J., 1979. Theory of population genetics and 
evolutionary ecology: an introduction. MacMillan, New York.

\noindent Sala, E., Graham, M.~H., 2002. Community-wide distribution of 
predator-prey interaction strength in kelp forests. Proc. Natl. Acad. 
Sci. USA 99, 3678-3683. 

\noindent Tavares-Cromar, A.~F., Williams, D.~D., 1996. 
The importance of temporal resolution in food web analysis: evidence from a detritus-based stream.
Ecol. Monogr. 66, 91-113.

\noindent Townsend, C.~R., Thompson, R.~M., McIntosh, A.~R., Kilroy, C., 
Edwards, E., Scarsbrook, M.~R., 1998. Disturbance, resource supply, 
and food-web architecture in streams. Ecol. Lett 1, 200-209.

\noindent Waide, R.~B., Reagan, D.~P., 1996. The food web of a tropical 
rain forest. University of Chicago Press, Chicago.

\noindent Waide, R.~B., Willig, M.~R., Steiner, C.~F., Mittelbach, G., Gough, L., Dodson, S.~I., Juday, G.~P., Parmenter, R., 1999.
The relationship between productivity and species richness. Annu. Rev. Ecol. Syst. 30, 257-300.

\noindent Walker, B.~H., 1992. Biodiversity and ecological redundancy. 
Conserv. Biol. 6, 18-23.

\noindent Warren, P.~H., 1989. Spatial and temporal variation in the 
structure of a fresh-water food web. Oikos 55, 299-311.

\noindent Williams, R.~J., Martinez, N.~D., 2000. Simple rules yield 
complex food webs. Nature 404, 180-183.

\noindent Williams, R.~J., Martinez, N.~D., 2004. Trophic levels in 
complex food webs: Theory and data. Am. Nat. in press.

\noindent Wootton, J.~T., 1997. Estimates and tests of per capita 
interaction strength: Diet, abundance, and impact of intertidally foraging 
birds. Ecol. Monogr. 67, 45-64.

\noindent Wright, D.~H., 1983. Species-energy theory: an extension of 
species-area theory. Oikos, 41, 496-506.

\noindent Yodzis, P., 1981. The stability of real ecosystems. Nature 289, 
674-676.

\end{document}